\documentclass[conference,compsoc]{IEEEtran}
\usepackage{amsmath,amssymb}
\usepackage{amsthm}
\usepackage{booktabs}
\usepackage{array}
\usepackage{xcolor}
\usepackage{url}
\usepackage{threeparttable}
\usepackage{graphicx}
\newtheorem{proposition}{Proposition}
\newtheorem{lemma}{Lemma}
\newtheorem{corollary}{Corollary}

\newcolumntype{L}[1]{>{\raggedright\arraybackslash}p{#1}}
\renewcommand{\paragraph}[1]{\smallskip\noindent\textbf{#1.}\hspace{0.5em}}
\ifCLASSOPTIONcompsoc
  \usepackage[nocompress]{cite}
\else
  \usepackage{cite}
\fi
\ifCLASSINFOpdf
\else
\fi
\begin{document}
%
\title{Erasing the Wrong Thing: Conditional Attribute Leakage and Concept Erasure in Shared Foundation-Model Embeddings}

\author{
\IEEEauthorblockN{
Haijun He\textsuperscript{1,2} \quad
Fei Li\textsuperscript{1} \quad
Donghong Ji\textsuperscript{1} \quad
Junjun He\textsuperscript{2}
}
\IEEEauthorblockA{
\textsuperscript{1}School of Cyber Science and Engineering, Wuhan University
\quad
\textsuperscript{2}Shanghai Innovation Institute\\
\{haijunhe,~lifei\_csnlp,~dhji\}@whu.edu.cn
\quad hejunjun@sjtu.edu.cn
}
}

\maketitle

\begin{abstract}
Medical-image embeddings are increasingly shared in place of raw images for authorized downstream tasks, yet they may reveal sensitive attributes. 
Standard concept-erasure methods target population-level guardedness. 
However, we prove that, when binary task and sensitive labels are associated, global and task-conditional linear guardedness cannot both hold while a nonzero linear task signal is preserved, so global erasure must either eliminate that signal or retain task-conditional linear leakage.
We further show how global erasure affects linear readers' discrimination between positive and negative samples with different sensitive attributes, deriving an approximate closed-form prediction of its utility cost from pre-erasure statistics.
Motivated by this conflict, we introduce Task-Conditional LEACE (TC-LEACE), a closed-form variant of LEACE that enforces linear guardedness within each task stratum of its fitting data. 
We evaluate four vision foundation models on 85 CelebA attribute pairs spanning a broad range of associations. 
After LEACE, conditional leakage persists across all four encoders, the discrepancy between marginal and conditional audits increases with association strength, and predicted utility costs track observed losses. 
We then examine hospital chest radiographs using three foundation models, including two biomedical models, in which chest-tube status is strongly associated with diagnosis. 
Although a marginal attack on LEACE-erased embeddings underperforms a diagnosis-only reference, chest-tube status remains detectable within diagnosis strata. 
TC-LEACE recovers most of LEACE's linear task-utility loss while keeping task-conditional leakage within a prespecified margin of LEACE's.
\end{abstract}


%
\IEEEpeerreviewmaketitle

\section{Introduction}
\label{sec:intro}

With the rapid development of medical imaging foundation models, precomputed image embeddings are becoming an increasingly attractive data-sharing format in multicenter medical research, since they can reduce storage and communication costs as well as the computational burden of downstream model training compared with sharing raw medical images~\cite{sellergren2023mimicembeddings,cong2026flexible,bian2026fully,array2026fmfl}.
However, prior studies have shown that medical imaging embeddings are not inherently privacy-preserving and may enable sensitive attribute inference beyond the intended purpose of data sharing~\cite{santos2025persistent,bajar2026brainmri,quarta2026hidden,yang2025demographic,kim2026encoding}.
This raises a question closely related to the \textit{Principle of Least Privilege}:
\textit{How can we restrict the inference capabilities exposed by shared embeddings beyond those required for an authorized task?}~\cite{saltzer1975protection,stadler2024leastprivilege}

\textbf{Concept erasure} offers a natural pre-release defense against such unintended inference capabilities. 
Post-hoc methods apply this defense to existing embeddings without retraining the foundation model. 
Among them, LEACE~\cite{belrose2023leace} ensures that, on the data used to fit the eraser, the transformed representations have identical means across sensitive attribute groups, a guarantee known as linear guardedness.
Because LEACE does not use the authorized task label, it constrains the linear predictability of sensitive attributes over the entire population, and we refer to this setting as \textbf{global concept erasure}.
However, suppressing and auditing sensitive attribute information at the population level may not match the privacy requirements of embedding sharing for a specific authorized task~\cite{eid2026leapsc}.

To illustrate this mismatch, consider an authorized task label $W$ and an unauthorized sensitive attribute $Z$ that are statistically associated.
Sensitive attribute inference from the shared embedding $E$ can arise from two distinct sources:
\[
\underbrace{Z\not\!\perp W}_{\text{Imputation}},
\qquad
\underbrace{E\not\!\perp Z\mid W}_{\text{Conditional Leakage}}.
\]
The first, \textbf{imputation}, arises from the statistical association between $W$ and $Z$, whereby $W$ may reveal information about $Z$ even without $E$~\cite{jayaraman2022imputation,stadler2024leastprivilege}.
The second, which we refer to as \textbf{conditional leakage}, concerns the additional information about $Z$ that $E$ provides beyond what $W$ reveals. 
Here, $W$ is used to define and evaluate this additional disclosure and need not be available to the data recipient.

This distinction reveals two potential failure modes of global concept erasure. 
\textbf{First, excessive erasure may impair authorized task utility.} 
When $W$ and $Z$ are associated, global erasure may remove directions that predict $W$, degrading task utility without concealing what $W$ already reveals about $Z$.
\textbf{Second, insufficient erasure may leave residual conditional leakage.} 
Even when the global linear association between the representation and $Z$ has been eliminated, information about $Z$ may remain linearly decodable among samples sharing the same task label $W$.

Existing concept erasure methods, including LEACE, primarily target population-level associations with sensitive attributes~\cite{belrose2023leace,chowdhury2025fundamental,holstege2025splince,akbari2026obliviator} and may therefore exhibit both failure modes described above.
Task-conditional constraints have been explored in fair representation learning, but primarily during representation training rather than post   hoc~\cite{zhang2018adversarial,zhao2020cfair,adeli2021representation, eid2026leapsc}.
In contrast, embedding sharing in multi-center medical research requires task-conditional erasure applied post hoc to precomputed embeddings.

To meet this requirement for shared embeddings, we analyze the utility loss and residual leakage that global concept erasure causes and build a task-conditional eraser.
\textbf{First}, we establish a task-conditional evaluation framework that distinguishes imputation from conditional leakage, revealing the limitations
of population-level privacy audits.
\textbf{Second}, we examine the relationship between global and task-conditional linear guardedness and show their incompatibility for any release that keeps its linear task signal whenever the task and the attribute are associated.
For linear readers, we further derive a closed-form prediction of the utility cost of global erasure from pre-erasure statistics and characterize when linear conditional leakage survives it.
\textbf{Third}, building on these findings, we introduce TC-LEACE, a closed-form variant of LEACE that applies a unified post-hoc projection to remove sensitive attribute mean differences within each task stratum on the eraser-fitting data, thereby satisfying task-conditional linear guardedness.

We test these findings on four general-purpose vision FMs across 85 CelebA~\cite{liu2015celeba} attribute pairs that span a broad range of associations, and on three FMs, including two biomedical encoders, on the CANDID-PTX chest radiograph dataset~\cite{feng2021candid}, using identity-level and patient-level splits, respectively.
Experiments show that conditional leakage persists in every encoder after LEACE, the gap between population-level and conditional audits widens as the association strengthens, and the predicted utility cost tracks the measured loss across CelebA pairs. 
In the medical case study, TC-LEACE recovers most of the linear task-utility loss of LEACE while keeping conditional leakage within a prespecified margin of that of LEACE, and on CelebA its gain concentrates on strongly associated attribute pairs.

\section{Background}
\label{sec:background}

\subsection{Sharing Foundation-Model Embeddings}
\label{sec:bg-sharing}
A foundation model maps an input image $X$ to an embedding $E=f(X)\in\mathbb{R}^d$ that can be reused across multiple downstream tasks.
For an authorized downstream task, a data custodian may share precomputed embeddings instead of raw images, with or without task labels depending on the sharing protocol.
Authorized recipients then train lightweight models, such as linear probes or small multilayer perceptrons (MLPs), on the frozen embeddings.
Practical uses of such representations include public releases of chest radiograph embeddings~\cite{sellergren2023mimicembeddings} and federated learning on frozen FM embeddings~\cite{array2026fmfl,cong2026flexible,bian2026fully}.
To restrict sensitive attribute inference while preserving utility for the authorized task, the custodian can apply a post-hoc transformation $r:\mathbb{R}^d\to\mathbb{R}^d$, fitted once on its own labeled data, to every embedding before release.
We denote the resulting representation by $E'=r(E)$, with $E'=E$ when no erasure is applied.

\subsection{Concept Erasure and Linear Guardedness}
\label{sec:bg-erasure}
Concept erasure suppresses information about a concept $Z$ in a representation while preserving the rest of the representation as much as possible.
Throughout, $Z$ is binary and both of its groups have positive probability.
The central notion in linear concept erasure is \emph{linear guardedness}.
A representation $E'$ linearly guards $Z$ if no affine predictor using $E'$ achieves a lower expected loss than the best constant predictor under any convex loss, subject to standard integrability conditions~\cite{ravfogel2023loglinear,belrose2023leace}.
This holds if and only if the class-conditional means are equal~\cite{belrose2023leace}:
\begin{equation}
\mathbb{E}[E' \mid Z=1]
=
\mathbb{E}[E' \mid Z=0]
\quad\Longleftrightarrow\quad
\mathrm{Cov}(E',Z)=0.
\label{eq:guard}
\end{equation}
Consequently, every affine score $s(E')=a^\top E'+c$ has the same mean in both groups.
This property forms the basis of our analysis in Section~\ref{sec:mechanism}.

LEACE (LEAst-squares Concept Erasure) attains linear guardedness while minimizing the expected squared change to the representation among affine erasers~\cite{belrose2023leace}, and we use it as our main global eraser.
Let $\mu=\mathbb{E}[E]$, $\Sigma=\mathrm{Cov}(E)$, and $\Sigma_{EZ}=\mathrm{Cov}(E,Z)$.
The LEACE transformation can be written as
\begin{equation}
r_{\mathrm{LEACE}}(E)
=
E-\Sigma^{1/2}P\Sigma^{+1/2}(E-\mu),
\label{eq:leace}
\end{equation}
where $\Sigma^{+1/2}:=(\Sigma^{1/2})^+$ denotes the Moore--Penrose pseudoinverse of $\Sigma^{1/2}$, which equals $\Sigma^{-1/2}$ when $\Sigma$ is invertible, and $P$ is the orthogonal projection onto the column space of $\Sigma^{+1/2}\Sigma_{EZ}$.
This column space has dimension one, or zero when $\Sigma_{EZ}=0$, so LEACE removes at most one direction in whitened coordinates.
In practice, the moments are estimated from the eraser-fitting data, so guardedness holds on that sample but not automatically on unseen data.

Other erasers impose different global constraints.
SPLINCE keeps linear guardedness while preserving the covariance between the representation and the task label~\cite{holstege2025splince}.
Distributional erasure methods seek independence between the representation and the sensitive attribute~\cite{chowdhury2025fundamental,akbari2026obliviator,saillenfest2025density,simai2026principled}.
Post-hoc orthogonalization regresses the embeddings on protected attributes and removes the fitted component, which requires these attributes for every released embedding~\cite{weber2025orthogonalization}.
Section~\ref{sec:mechanism} analyzes SPLINCE and distributional erasure, and Section~\ref{sec:case} evaluates orthogonalization.

Linear guardedness does not imply $E'\perp Z$, so nonlinear predictors such as MLPs may still recover $Z$ from $E'$~\cite{elazar2018adversarial,ravfogel2020inlp,kumar2022probing,ravfogel2023loglinear}, and our evaluation therefore includes nonlinear attackers.
Equality of group means also does not make every linear score uninformative under AUROC, which measures pairwise ranking rather than expected convex loss.
However, a linear probe fitted to optimality on the fitting data with a strictly convex loss, such as the logistic loss, outputs a constant score and therefore attains an AUROC of exactly $0.5$ on that data by construction.
Because none of these constraints conditions on the authorized task label, we refer to Eq.~\eqref{eq:guard} as \emph{global} linear guardedness, in contrast to the task-conditional constraint that Section~\ref{sec:method} imposes within each stratum of the task.

\subsection{AUROC and Its Pairwise Decomposition}
\label{sec:bg-auroc}
We evaluate both authorized task utility and sensitive attribute inference using the area under the receiver operating characteristic curve (AUROC).
Let $R$ denote the representation under evaluation, either $E$ or $E'$.
For a scoring function $s$ and a binary label $Y$, we denote the AUROC by $\mathrm{AUC}(s;Y)$ and define it as
\begin{equation}
\mathrm{AUC}(s;Y)
=
\Pr\big[s(R_+)>s(R_-)\big]
+
\frac{1}{2}
\Pr\big[s(R_+)=s(R_-)\big],
\label{eq:auc}
\end{equation}
where $\Pr[\cdot]$ denotes probability, and $R_+$ and $R_-$ are drawn independently from the positive ($Y=1$) and negative ($Y=0$) populations, respectively.

AUROC is threshold-independent and invariant under strictly increasing transformations of the score.
A score independent of the label has an expected AUROC of $0.5$, irrespective of class prevalence.
These properties make AUROC useful for comparing ranking performance across attributes with different prevalences.
However, an overall AUROC of $0.5$ can hide discrimination within subgroups, which the following decomposition makes explicit.

Because AUROC is defined over positive--negative sample pairs, it admits an exact decomposition when samples are additionally partitioned by a binary grouping variable $G$.
Let $\mathrm{AUC}_{gh}$ denote Eq.~\eqref{eq:auc} with $R_+$ drawn from positive samples with $G=g$ and $R_-$ drawn from negative samples with $G=h$, using the same scoring function $s$.
By the law of total probability,
\begin{equation}
\mathrm{AUC}(s; Y) = \sum_{g,h \in \{0,1\}} \omega_{gh} \, \mathrm{AUC}_{gh},
\label{eq:decomp}
\end{equation}
where $\omega_{gh} = \Pr[G=g \mid Y=1]\,\Pr[G=h \mid Y=0]$ and the sum runs over combinations with $\omega_{gh}>0$.
Terms with $g=h$ compare positive and negative samples within the same group, whereas terms with $g\neq h$ compare samples across different groups.
Such cross-group comparisons are related to the xAUC fairness metric of Kallus and Zhou~\cite{kallus2019xauc} and to other pairwise fairness criteria~\cite{beutel2019pairwise,vogel2021fairscoring}.

We use Eq.~\eqref{eq:decomp} in two ways.
First, setting $Y=W$ and $G=Z$, we compare the pairwise components before and after erasure to identify which sample pairs account for the loss of authorized task utility (Section~\ref{sec:mechanism}).
Second, setting $Y=Z$ and $G=W$, we decompose sensitive attribute inference according to the task labels of the positive and negative samples.
The within-group terms $\mathrm{AUC}_{00}$ and $\mathrm{AUC}_{11}$ then measure sensitive attribute discrimination within each task stratum, provided both classes of $Z$ occur in each stratum.
Eq.~\eqref{eq:decomp} links these terms to the overall AUROC only when both are computed for the same score, which may itself depend on $W$.
Section~\ref{sec:defs} defines conditional AUROC by renormalizing these within-stratum terms.

\begin{table}[t]
\centering
\caption{Parties in authorized embedding sharing.}
\label{tab:threat}
\small
\begin{tabular}{@{}L{0.17\columnwidth}L{0.39\columnwidth}L{0.32\columnwidth}@{}}
\toprule
Party & Access & Objective \\
\midrule
Custodian & Raw data, labels of $W$ and $Z$, public FM & Release $E'$ that supports $W$ and limits disclosure of $Z$ beyond $W$ \\
Honest recipient & $E'$ and $W$ for training samples & Train a linear or nonlinear model for $W$ \\
Attacker & $E'$, $Z$ for a subset of released samples, and $W$ where released or under conditional evaluation & Infer $Z$ for the remaining released samples \\
\bottomrule
\end{tabular}
\end{table}

\section{Problem Formulation}
\label{sec:setting}

\subsection{Setting and Threat Model}
\label{sec:threat}
A \emph{\textbf{data custodian}} holds raw data with labels for an authorized task $W$ and a sensitive attribute $Z$, and has access to a public foundation model.
It computes embeddings, fits an eraser on its own data, and releases the erased embeddings $E'$, aiming to keep them useful for $W$ while limiting what they disclose about $Z$ beyond $W$.
Fitting any eraser requires labels of $Z$, and the task-conditional eraser of Section~\ref{sec:method} requires joint labels of $W$ and $Z$.
Labels of $W$ typically accompany downstream training samples, whereas for new samples $W$ is what recipients are authorized to predict.
An \emph{\textbf{honest recipient}} uses the release to train a linear or nonlinear model for $W$.
An \emph{\textbf{attacker}}, who may be a curious recipient or any other party that obtains the release, aims to infer $Z$ with models of any class and may use $W$ wherever it is released.
The erasers we study guarantee protection only against linear attackers on their fitting data, so we evaluate nonlinear attackers empirically to measure what this guarantee leaves open.
We assume that the attacker knows $Z$ for some released samples, so it can train on erased embeddings without access to the eraser.
Table~\ref{tab:threat} summarizes each party's information and objectives.

In our medical case study, the custodian releases chest radiograph embeddings for detecting pneumothorax ($W$).
The sensitive attribute $Z$ is the presence of a chest tube, which reveals past invasive treatment beyond the authorized diagnosis.
An attacker could be a participating site that uses clinical records of its own patients to infer $Z$ for other patients in the release.

We consider two cases for the attacker's access to $W$.
In conditional evaluation, the attacker observes the true task label with $E'$, regardless of whether $W$ accompanies the release in deployment.
This case gives the attacker at least as much information as any release protocol, so it does not underestimate leakage and isolates what $E'$ adds beyond the authorized task.
In unconditional evaluation, the attacker observes only $E'$.
This case reflects the common practice of auditing erased embeddings alone, but it understates the attacker's capability when the release also includes $W$.

\begin{figure*}[t]
\centering
\includegraphics[width=\textwidth]{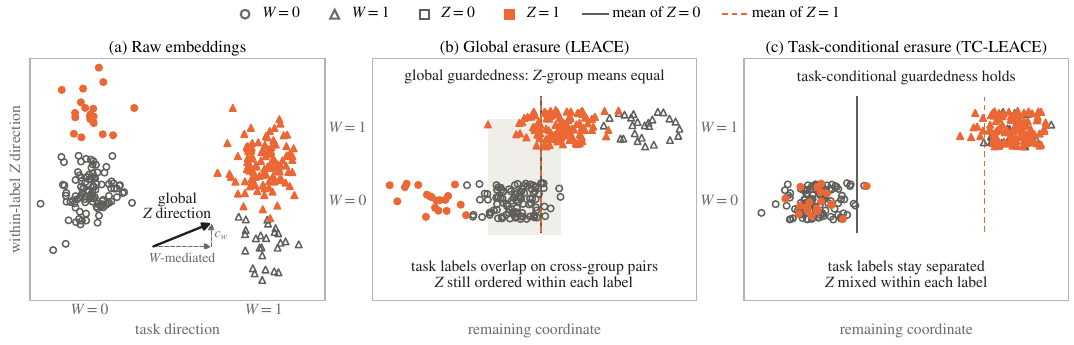}
\caption{Two-dimensional toy example of global and task-conditional erasure, with $W$ associated with $Z$.
(a)~The global $Z$ direction combines a $W$-mediated component along the task direction with a within-label component built from $c_0$ and $c_1$ (Eq.~\eqref{eq:covdecomp}), and the arrows indicate directions only.
(b)~LEACE removes the global direction, so the $Z$ groups have equal means and a fitted linear probe for $Z$ is constant.
$Z$ remains ordered within each label, and the task labels overlap (shaded) only where samples with $W=1, Z=1$ meet samples with $W=0, Z=0$, the cross-group pairs of Section~\ref{sec:fail1}.
(c)~Task-conditional erasure removes only the within-label direction, so the task labels stay separated and $Z$ is inferable only through $W$.
Panels~(b) and~(c) plot the single remaining coordinate, with one row per task label and vertical jitter.}
\label{fig:concept}
\end{figure*}

\subsection{Permitted Inference and Conditional Leakage}
\label{sec:defs}
Throughout, $W$ and $Z$ are binary, and all four combinations of their values have positive probability.
We write
\[
\pi_z = \Pr[W=1 \mid Z=z], \qquad \rho_w = \Pr[Z=1 \mid W=w],
\]
and we quantify the signed association between the authorized task and the sensitive attribute by $\beta = \pi_1 - \pi_0$.
With $p = \Pr[Z=1]$ and $q = \Pr[W=1]$,
\[
\mathrm{Cov}(W,Z) = p(1-p)\,\beta = q(1-q)\,(\rho_1-\rho_0),
\]
so $\beta$ and $\rho_1 - \rho_0$ share the sign of $\mathrm{Cov}(W,Z)$, and both vanish exactly when $W$ and $Z$ are independent.

\paragraph{Imputation}
Any party that knows $W$ can score samples by $W$, or by $1-W$ when $\beta<0$.
Using these scores in Eq.~\eqref{eq:auc} gives
\begin{equation}
\mathrm{AUC}(W;Z)=\frac12+\frac{\beta}{2},
\qquad
\mathrm{AUC}_{\mathrm{imp}}^*=\frac12+\frac{|\beta|}{2}.
\label{eq:ceiling}
\end{equation}
The quantity $\mathrm{AUC}_{\mathrm{imp}}^*$ is the maximum AUROC achievable from $W$ alone, which we call the \emph{imputation ceiling}.
This inference requires no access to embeddings and corresponds to the imputation of Jayaraman and Evans~\cite{jayaraman2022imputation} and the fundamental leakage of Stadler et al.~\cite{stadler2024leastprivilege}, instantiated for a binary task and measured by AUROC.

In general, the ceiling does not bound attackers that use embeddings.
It does bound them when the embeddings disclose nothing beyond the task.
If $E' \perp Z \mid W$, the distribution of $E'$ given $Z$ is a mixture of its distributions given $W=0$ and $W=1$, with weights $\Pr[W=w \mid Z]$.
With respect to $Z$, the embedding is then a randomized function of $W$, and no score computed from $E'$, with or without $W$, can exceed the ceiling (Appendix~\ref{app:proofs}).
An AUROC above the ceiling therefore reveals conditional leakage, whereas an AUROC at or below it does not rule it out.

\paragraph{Conditional leakage}
Disclosure beyond the authorized task is the dependence between $Z$ and $E'$ that remains among samples with the same value of $W$, that is, $E' \not\perp Z \mid W$.
Figure~\ref{fig:concept}(b) illustrates it: after global erasure the two $Z$ groups have equal means, yet $Z$ remains ordered within each task label.
We assess it empirically with the \emph{conditional AUROC} of an attacker score $s(E',W)$, which may treat the two strata differently:
\begin{equation}
\begin{aligned}
\mathrm{AUC}_{\mathrm{cond}}(s)
&= \sum_{w\in\{0,1\}} v_w\,
   \mathrm{AUC}(s;Z\mid W=w),\\
v_w
&= \frac{n_{w1}n_{w0}}
{\sum_{u\in\{0,1\}}n_{u1}n_{u0}},
\end{aligned}
\label{eq:condauc}
\end{equation}
where $\mathrm{AUC}(s;Z\mid W=w)$ denotes Eq.~\eqref{eq:auc} evaluated on samples with $W=w$, and $n_{wz}$ denotes the number of evaluation samples with $W=w$ and $Z=z$.
With $Y=Z$ and $G=W$ in Eq.~\eqref{eq:decomp}, the conditional AUROC corresponds to the within-group terms after renormalization.

We use $\mathrm{AUC}_{\mathrm{cond}}(s)-0.5$ as an empirical measure of conditional leakage, the signed advantage of the evaluated attacker over chance within strata.
Because $E' \perp Z \mid W$ implies an AUROC of $0.5$ within each stratum for every score, a value of zero is necessary but not sufficient for conditional independence.
It can arise from an attacker that fails to exploit the dependence, or from a weighted average that hides a stratum where the score performs well, so the AUROC of each stratum should also be examined.

\paragraph{Unconditional inference}
The common measure of attribute inference is the unconditional AUROC $\mathrm{AUC}(s;Z)$ of an attacker that observes only $E'$.
With $Y=Z$ and $G=W$ in Eq.~\eqref{eq:decomp}, let $c=\omega_{10}+\omega_{01}$ denote the proportion of positive--negative pairs drawn from different task-label strata.
We define the cross-stratum AUROC as
\[
\mathrm{AUC}_{\times}(s)
=
\frac{
\omega_{10}\mathrm{AUC}_{10}(s)
+
\omega_{01}\mathrm{AUC}_{01}(s)
}{c}.
\]
For a fixed score $s(E')$, evaluated both unconditionally and within each stratum, the decomposition gives
\begin{equation}
\begin{aligned}
&\mathrm{AUC}(s;Z)-\mathrm{AUC}_{\mathrm{cond}}(s)\\
&\qquad =
c\big(
\mathrm{AUC}_{\times}(s)
-
\mathrm{AUC}_{\mathrm{cond}}(s)
\big).
\end{aligned}
\label{eq:gap}
\end{equation}
Thus, the difference between unconditional and conditional AUROC for the same score is determined by its cross-stratum ranking performance.

The relative weights of the two cross-stratum pair types depend on the association between $W$ and $Z$, since $\omega_{10}-\omega_{01}=\pi_1(1-\pi_0)-(1-\pi_1)\pi_0=\beta$.
A score based solely on $W$, oriented according to the sign of $\beta$, exploits this imbalance and attains the imputation ceiling, an advantage of $|\beta|/2$ over chance.
More generally, an embedding-based attacker may exploit the association between $W$ and $Z$ without discriminating $Z$ within each task-label stratum.
Moreover, since $c \ge |\omega_{10}-\omega_{01}| = |\beta|$, pairs within strata carry a weight of at most $1-|\beta|$ in the unconditional AUROC, so for a strongly associated task the unconditional AUROC is driven mainly by cross-stratum ranking.

Neither measure bounds the other in general.
A score based solely on $W$ reaches the imputation ceiling unconditionally but has a conditional AUROC of $0.5$.
Conversely, by Eq.~\eqref{eq:gap}, the unconditional AUROC falls below the conditional AUROC whenever the score ranks cross-stratum pairs worse than within-stratum pairs, as in Figure~\ref{fig:concept}(b).
This can happen even when $\beta = 0$, for example when the level of the score differs between the two strata.

\subsection{Limitations of Population-Level Audits}
\label{sec:audits}
We examine two common population-level audits, summarized in Table~\ref{tab:audits}.

\paragraph{Probe-to-chance}
The custodian trains a linear probe for $Z$ on the erased embeddings and checks that its AUROC is close to $0.5$.
On the data used to fit LEACE, this outcome holds by construction (Section~\ref{sec:bg-erasure}), so it provides no evidence about leakage on new data.
A held-out probe measures the linear predictability of $Z$ for the tested probe but does not establish guardedness on new data, and neither probe assesses nonlinear or task-conditional inference.

\paragraph{Comparison with a label-only baseline}
The custodian compares the unconditional attack AUROC with what the task label alone achieves and concludes that no excess leakage exists when the attack does not exceed this baseline.
Dataset leakage~\cite{wang2019balanced} and the label-only floor~\cite{eid2026leapsc} apply the same idea with accuracy-based measures.
By Section~\ref{sec:defs}, this comparison is one-sided.
Erasure may degrade the cross-stratum ranking of an attacker without $W$ and so lower its unconditional AUROC below the ceiling, while its within-stratum ranking, which carries little weight when $|\beta|$ is large, still separates $Z$.

Both audits measure inference over the whole population, so they can miss residual inference within the strata of the authorized task and conflate task-mediated inference with disclosure by the embeddings.
Section~\ref{sec:case} shows a release that passes the label-only comparison while disclosing $Z$ within strata.

\begin{table}[t]
\centering
\caption{What each leakage evaluation measures.}
\label{tab:audits}
\small
\begin{tabular}{@{}L{0.27\columnwidth}L{0.29\columnwidth}L{0.32\columnwidth}@{}}
\toprule
Evaluation & Measures & Does not measure \\
\midrule
In-sample linear probe & Moment constraint of the eraser on the fitting data & Leakage on new data, nonlinear and within-stratum inference \\
Held-out linear probe & Linear predictability of $Z$ for the tested probe & Nonlinear and within-stratum inference \\
Unconditional attack & Inferability of $Z$ from $E'$ alone & Separation of imputation from leakage \\
Unconditional attack vs.\ label-only baseline & Conditional leakage, if the attack exceeds the ceiling & Absence of conditional leakage \\
Conditional attack (ours) & Within-stratum inference for the tested attackers & Attackers outside the tested class \\
\bottomrule
\end{tabular}
\end{table}
\section{Why Global Erasure Fails}
\label{sec:mechanism}
Section~\ref{sec:setting} judges a release by what it discloses beyond the authorized task, but global erasure constrains a different quantity.
We analyze the two resulting failure modes, lost authorized task utility and residual within-stratum information about $Z$ (Figure~\ref{fig:concept}), and Section~\ref{sec:readers} examines their dependence on the reader class and on the choice of global eraser.
Appendix~\ref{app:proofs} contains the proofs.

\subsection{Failure 1: Guardedness Spends Utility on Permitted Inference}
\label{sec:fail1}
Consider an affine task score $s(E') = a^\top E' + b$, and for $z \in \{0,1\}$ let $m_z = \mathbb{E}[s \mid W=0, Z=z]$ and $d_z = \mathbb{E}[s \mid W=1, Z=z] - m_z$, both in the units of $s$.
The mean score of group $Z=z$ is $m_z + \pi_z d_z$.
Because linear guardedness equalizes these means (Section~\ref{sec:bg-erasure}),
\begin{equation}
m_1 - m_0 = \pi_0 d_0 - \pi_1 d_1
\label{eq:shift}
\end{equation}
holds on fitting data without distributional assumptions.

\begin{proposition}[Guardedness excludes informative leakage-free task scores]
\label{prop:means}
Let $\beta \neq 0$. (a)~If $E'$ linearly guards $Z$, no affine score $s$ of $E'$ satisfies $\mathbb{E}[s \mid W,Z] = \mathbb{E}[s \mid W]$ and $\mathbb{E}[s \mid W=1] \neq \mathbb{E}[s \mid W=0]$. (b)~If $E' \perp Z$, no measurable score $s$ of $E'$ satisfies $s \perp Z \mid W$ and $s \not\perp W$.
\end{proposition}

Part~(b) is the incompatibility between independence and separation known from fair classification~\cite{barocas2023fairmlbook}, and part~(a) is its linear counterpart.
A score whose mean discloses nothing about $Z$ beyond $W$ differs between the groups of $Z$ by $\beta(\mathbb{E}[s \mid W=1] - \mathbb{E}[s \mid W=0])$, which is the association that $W$ itself discloses.
Global guardedness forbids exactly this difference, so every informative linear task score must cancel it, either by shifting between the groups of $Z$ or by separating the task classes differently in them.

When $d_0 = d_1 = d$, Eq.~\eqref{eq:shift} requires $m_1 - m_0 = -\beta d$.
In an idealized model where this offset is the only change to the score, it acts only on the cross-group pairs of Eq.~\eqref{eq:decomp} with $Y=W$ and $G=Z$.
In Figure~\ref{fig:concept}(b), the task labels overlap only where samples with $W=1, Z=1$ meet samples with $W=0, Z=0$.
Actual erasers can also change within-group rankings, and Sections~\ref{sec:case} and~\ref{sec:scale} measure both parts.

\begin{corollary}[Cross-group AUROC under guardedness]
\label{cor:xauc}
If, in addition, $s$ is normal within each $(W,Z)$ group with a common variance $\sigma^2 > 0$ across the four groups, then with $\Delta_z = d_z/\sigma = \sqrt{2}\,\Phi^{-1}(\mathrm{AUC}_{zz})$ and $\delta = (m_1 - m_0)/\sigma = \pi_0\Delta_0 - \pi_1\Delta_1$,
\begin{equation}
\mathrm{AUC}_{10} = \Phi\big(\tfrac{\Delta_1 + \delta}{\sqrt{2}}\big),
\quad
\mathrm{AUC}_{01} = \Phi\big(\tfrac{\Delta_0 - \delta}{\sqrt{2}}\big),
\label{eq:xauc}
\end{equation}
where $\mathrm{AUC}_{10}$ pairs task positives with $Z=1$ against task negatives with $Z=0$, and $\mathrm{AUC}_{01}$ the reverse.
\end{corollary}

When the within-group separations are equal, $\Delta_0 = \Delta_1 = \Delta$, Eq.~\eqref{eq:xauc} reduces to $\mathrm{AUC}_{10} = \Phi\big(\tfrac{(1-\beta)\Delta}{\sqrt{2}}\big)$ and $\mathrm{AUC}_{01} = \Phi\big(\tfrac{(1+\beta)\Delta}{\sqrt{2}}\big)$, whereas both would equal the within-group value $\Phi\big(\tfrac{\Delta}{\sqrt{2}}\big)$ without the offset.
For $\beta > 0$, the pairs that follow the association lose separation in proportion to $\beta$ and approach chance as $\beta \to 1$, while the reverse pairs gain.
On new data, where guardedness is approximate, the same derivation gives $\delta = \pi_0\Delta_0 - \pi_1\Delta_1 + \varepsilon$ with $\varepsilon = (\mathbb{E}[s \mid Z=1] - \mathbb{E}[s \mid Z=0])/\sigma$ on the same data, so inserting a measured $\varepsilon$ is a post-hoc adjustment.
For an ex-ante estimate, we set $\varepsilon = 0$ and apply the corollary to the within-group AUROCs of the raw embeddings, assuming erasure preserves them.
Eq.~\eqref{eq:decomp} turns the predicted cross-group AUROCs into an estimate of the cross-group loss under that assumption.
Section~\ref{sec:scale} tests this estimate across attribute pairs, and Appendix~\ref{app:more} checks it in simulation.

How much this loss matters depends on the cross-group pair weights, which for $Y=W$ and $G=Z$ are $\omega_{10} = \rho_1(1-\rho_0)$ and $\omega_{01} = (1-\rho_1)\rho_0$, with difference $\rho_1 - \rho_0$.
The pairs that follow the association (for $\beta > 0$, task positives with $Z=1$ against task negatives with $Z=0$) therefore always form the larger cross-group type.
For $\Delta > 0$, they are also the pairs whose AUROC the offset lowers.
The net change in task AUROC also depends on the size of these effects and, for actual erasers, on changes within groups.

\subsection{Failure 2: Within-Stratum Signal Survives Global Erasure}
\label{sec:fail2}
Let $c_w = \mathrm{Cov}(E, Z \mid W=w)$ denote the within-stratum cross-covariance and $\mu_w = \mathbb{E}[E \mid W=w]$.
By the law of total covariance,
\begin{equation}
\Sigma_{EZ} = (1-q)\,c_0 + q\,c_1
+ q(1-q)\,(\mu_1 - \mu_0)(\rho_1 - \rho_0).
\label{eq:covdecomp}
\end{equation}
We call the last term the $W$-mediated component.
It is large when the task-conditional means $\mu_0$ and $\mu_1$ differ substantially and $W$ is strongly associated with $Z$.
Figure~\ref{fig:concept}(a) shows the two components, with the $W$-mediated component along the task direction and the within-stratum component along the within-label direction.

Eq.~\eqref{eq:covdecomp} also holds for $E'$.
Global guardedness zeroes its left side, so with primes denoting the moments of $E'$,
\begin{equation}
(1-q)\,c_0' + q\,c_1' = -q(1-q)\,(\rho_1 - \rho_0)(\mu_1' - \mu_0').
\label{eq:guarddecomp}
\end{equation}
When $\beta \neq 0$, a guarded release whose task classes still differ in mean ($\mu_1' \neq \mu_0'$) therefore retains within-stratum covariance with $Z$ in at least one stratum.
This is part~(a) of Proposition~\ref{prop:means} stated for the embedding, and it makes global and task-conditional linear guardedness incompatible for any release that keeps its linear task signal.
Figure~\ref{fig:concept} illustrates this incompatibility, since panel~(b) satisfies global but not task-conditional guardedness, panel~(c) satisfies the reverse, and both keep the task signal.

\begin{lemma}[Residual within-stratum covariance]
\label{lem:residual}
Let $E' = r_{\mathrm{LEACE}}(E)$ and $M = I - \Sigma^{1/2} P \Sigma^{+1/2}$.
Then $\mathrm{Cov}(E', Z \mid W=w) = M c_w$, and $M c_w = 0$ if and only if $c_w \in \operatorname{span}(\Sigma_{EZ})$.
\end{lemma}

Since the kernel of $M$, $\operatorname{span}(\Sigma_{EZ})$, is at most one-dimensional, LEACE leaves within-stratum covariance in at least one stratum whenever $c_0$ and $c_1$ are nonzero and not parallel.
In whitened coordinates, the residual $(I - P)\,\Sigma^{+1/2} c_w$ is the within-stratum signal orthogonal to the erased direction, and when it is nonzero, $Z$ is not linearly guarded within stratum $w$.
This condition does not involve $\beta$, so global linear guardedness does not imply task-conditional linear guardedness even when $W$ and $Z$ are independent.

\subsection{Reader Classes and Other Global Erasers}
\label{sec:readers}
Part~(a) of Proposition~\ref{prop:means} and Eq.~\eqref{eq:guarddecomp} constrain only linear scores, exactly the readers that linear guardedness covers.
Nonlinear recipients may recover task information that remains nonlinearly accessible, although nothing guarantees it, and the guarantee does not cover nonlinear attackers either.
The cost and protection of a linear eraser thus share one scope, and Sections~\ref{sec:case} and~\ref{sec:scale} measure both.

Perfect global erasure enforces $E' \perp Z$, which leaves an attacker who observes $E'$ alone with only the prior of $Z$~\cite{chowdhury2025fundamental,akbari2026obliviator}.
By part~(b) of Proposition~\ref{prop:means}, when $\beta \neq 0$, a release that satisfies $E' \perp Z$ but retains information about $W$ must satisfy $E' \not\perp Z \mid W$, so perfect global erasure cannot remove conditional leakage from a useful release.
When $\beta = 0$, it does not exclude conditional leakage either.
If $W$ and $Z$ are independent fair bits and $E' = W \oplus Z$, then $E' \perp Z$, yet $E'$ reveals $Z$ once $W$ is known.

Preserving task covariance does not resolve the conflict.
SPLINCE enforces $\mathrm{Cov}(E', Z) = 0$ while keeping $\mathrm{Cov}(E', W) = \mathrm{Cov}(E, W)$, feasible when $\mathrm{Cov}(E, W)$ is not parallel to $\Sigma_{EZ}$~\cite{holstege2025splince}.
Since $\mathrm{Cov}(E', W) = q(1-q)(\mu_1' - \mu_0')$, SPLINCE keeps the linear task signal by design, and Eq.~\eqref{eq:guarddecomp} shows that when $\beta \neq 0$ it necessarily retains within-stratum covariance with $Z$.
For binary $Z$, SPLINCE and LEACE each remove one dimension and share the kernel $\operatorname{span}(\Sigma_{EZ})$, and such projections yield identical predictions for unregularized linear models retrained to optimality~\cite{holstege2025splince}.
The conflict stems from the global target, not from the eraser's construction.

\paragraph{Finite samples}
The guarantees above hold on the data used to fit an eraser and need not hold on new data, where estimation error can leave residual linear dependence and let a held-out probe exceed chance.
For a global signal that is strong relative to this error, the squared norm of the residual mean difference scales with $d/n$ in whitened coordinates rather than with the signal strength, where $n$ is the size of the fitting sample (Appendix~\ref{app:finite}).

\section{Task-Conditional Erasure}
\label{sec:method}
\begin{figure}[!htb]
\centering
\includegraphics[width=\columnwidth]{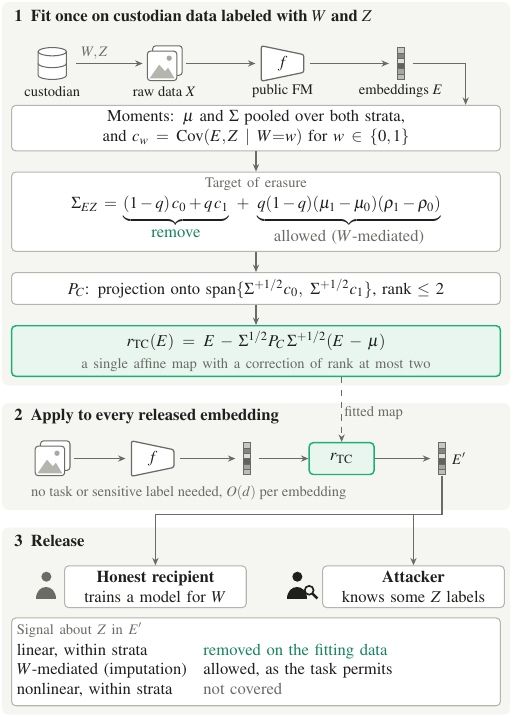}
\caption{The \textsc{TC-LEACE} pipeline.
The custodian fits the map once on data labeled with $W$ and $Z$, removing within-stratum cross-covariance and allowing the $W$-mediated component.
The same map is applied to every released embedding without labels.
Within strata, $E'$ carries no linear signal about $Z$ on fitting data, while nonlinear signal is not covered.}
\label{fig:pipeline}
\end{figure}

By Eq.~\eqref{eq:guarddecomp}, when $\beta \neq 0$, a release that keeps its linear task signal cannot be guarded both globally and within each stratum.
Only the conditional constraint matches the authorized task, so we change the target of erasure rather than its strength.
Figure~\ref{fig:pipeline} shows the three-stage procedure.

\paragraph{Fitting}
The custodian embeds its own data, which carry labels of $W$ and $Z$, and estimates the mean $\mu$ and covariance $\Sigma$ of $E$ pooled over both strata, together with the within-stratum cross-covariances $c_w = \mathrm{Cov}(E, Z \mid W=w)$ of Section~\ref{sec:fail2}.
Of the two components of $\Sigma_{EZ}$ in Eq.~\eqref{eq:covdecomp}, the eraser removes the within-stratum signals $c_0$ and $c_1$ and does not require the $W$-mediated component to vanish, which in Figure~\ref{fig:concept}(a) means removing the within-label direction and keeping the task direction.
Let $P_C$ be the orthogonal projection onto $\operatorname{span}\{\Sigma^{+1/2}c_0,\ \Sigma^{+1/2}c_1\}$, and define
\begin{equation}
r_{\mathrm{TC}}(E) = E - \Sigma^{1/2} P_C \, \Sigma^{+1/2}(E - \mu).
\label{eq:tc}
\end{equation}
We call this eraser Task-Conditional LEACE (\textsc{TC-LEACE}).
\textsc{TC-LEACE} is LEACE in Eq.~\eqref{eq:leace} applied to the two-dimensional concept $T=(T_0,T_1)$ with $T_w=\mathbb{1}\{W=w\}(Z-\rho_w)$, since $\operatorname{Cov}(E,T_w)=\Pr(W=w)\,c_w$.
It therefore inherits the least-squares optimality of LEACE among affine erasers that set both within-stratum cross-covariances to zero, and our contribution lies in this target and its analysis rather than in the solver.
The rank of $P_C$ equals the dimension of $\operatorname{span}\{c_0, c_1\}$, which is at most two for binary $W$ and $Z$, and dense fitting costs $O(nd^2 + d^3)$.

\paragraph{Deployment}
The result is a single affine map with a low-rank correction, which the custodian applies to every embedding before release.
With precomputed low-rank factors, this costs $O(d)$ per embedding and requires neither its task label nor retraining of the foundation model.
Label-specific projectors instead need the task label or its prediction to select a transformation at release~\cite{eid2026leapsc}, and representation-learning approaches to bias mitigation and conditional independence~\cite{adeli2021representation,hwa2024enforcing} require training an encoder or adapter.

\paragraph{What the release discloses}
The bottom of Figure~\ref{fig:pipeline} lists three kinds of signal about $Z$ in the released embedding $E' = r_{\mathrm{TC}}(E)$.
The first, linear signal within strata, is removed on the fitting data.

\begin{proposition}[Task-conditional linear guardedness]
\label{prop:tc}
On the distribution used to fit Eq.~\eqref{eq:tc}, $\mathrm{Cov}(r_{\mathrm{TC}}(E), Z \mid W=w) = 0$ for $w \in \{0,1\}$.
Equivalently, within each stratum of $W$, every affine score of the erased embedding has the same mean in both groups of $Z$.
\end{proposition}

Unlike under global guardedness (Proposition~\ref{prop:means}(a)), informative task scores carry no conditional mean leakage.
Eq.~\eqref{eq:shift} therefore no longer forces an offset between $Z$ groups, and the constraint causes no cross-group cost (Section~\ref{sec:fail1}).
Utility can still fall when the whitened task mean difference $\Sigma^{+1/2}(\mu_1-\mu_0)$ lies partly in the range of $P_C$.

The second, the $W$-mediated association, remains proportional to the linear task signal $\mathrm{Cov}(E', W)$.
Applying Eq.~\eqref{eq:covdecomp} to $E'$ with Proposition~\ref{prop:tc} and $\mathrm{Cov}(E', W) = q(1-q)(\mu_1' - \mu_0')$ gives
\begin{equation}
\mathrm{Cov}(E', Z) = (\rho_1 - \rho_0)\,\mathrm{Cov}(E', W),
\label{eq:tcglobal}
\end{equation}
which is the global covariance that $E' \perp Z \mid W$ would imply.
This covariance is nonzero whenever $\beta \neq 0$ and the linear task signal is kept, so \textsc{TC-LEACE} gives up global linear guardedness by design (Figure~\ref{fig:concept}(c)).

The third, nonlinear signal within strata, lies outside the guarantee.
\textsc{TC-LEACE} enforces only the linear, mean-level part of $E' \perp Z \mid W$, the representation-level analogue of equalized odds~\cite{hardt2016equality,zhang2018adversarial,zhao2020cfair}.
Sections~\ref{sec:case} and~\ref{sec:scale} also evaluate it against nonlinear attackers.

\section{Evaluation Methodology}
\label{sec:setup}

\subsection{Datasets and Foundation Models}
\label{sec:data}
Our questions place four requirements on a dataset.
(a)~Every sample needs labels of $W$ and $Z$, which \textsc{TC-LEACE} and the conditional AUROC use jointly.
(b)~$Z$ should be reflected in the raw data and reliably labeled, so that what an embedding encodes about it comes from the input itself.
(c)~$W$ and $Z$ should be associated but not perfectly, so that imputation and the cost of global erasure arise and every stratum of $W$ keeps both groups of $Z$.
(d)~Each sample should identify its source, so that all samples from one source fall on one side of every split.

To simulate a setting of a real custodian (Section~\ref{sec:case}), we need data with a genuinely sensitive attribute from a dataset that the encoders have not seen during pretraining.
Medical data are a typical instance of this setting, and chest radiographs in particular are covered by many medical foundation models.
However, widely used public chest radiograph datasets, such as ChestX-ray14~\cite{wang2017chestxray}, MIMIC-CXR~\cite{johnson2019mimiccxr}, and PadChest~\cite{bustos2020padchest}, derive their labels from radiology reports rather than expert image annotation, and all three were used to pretrain medical foundation models~\cite{perezgarcia2025raddino}, so they cannot stand in for a custodian's own data.
In contrast, CANDID-PTX meets these requirements.
Experts segmented its chest tubes~\cite{feng2021candid}, the dataset is available only under controlled access~\cite{candidptx_record}, and the documented pretraining data of the encoders we use, namely the medical encoders RAD-DINO~\cite{perezgarcia2025raddino} and BiomedCLIP~\cite{zhang2025biomedclip} and the general-purpose DINOv2~\cite{oquab2024dinov2}, do not include it.

To test how the effects vary with $|\beta|$ (Section~\ref{sec:scale}), we need many attribute pairs that span a broad range of associations, whereas a single medical pair fixes one value of $\beta$ and public medical datasets rarely provide reliable labels for many such pairs.
We therefore add CelebA, whose many binary attributes and identity labels meet this need.
Accordingly, we use general-purpose rather than medical encoders for it.

\paragraph{CANDID-PTX}
CANDID-PTX contains 19,237 chest radiographs from Dunedin Hospital, New Zealand, in which experts segmented pneumothoraces, acute rib fractures, and intercostal chest tubes, and it records a patient index for every image~\cite{feng2021candid}.
Given that a chest tube reveals treatment history, which is health data under the GDPR~\cite{gdpr2016} and health information under HIPAA~\cite{hipaa160103}, and that chest tubes are clinically associated with pneumothorax and a documented shortcut for pneumothorax classifiers~\cite{oakdenrayner2020hidden,rueckel2020impact}, we set $W$ to pneumothorax and $Z$ to the presence of a chest tube, as in Section~\ref{sec:threat}.
We fixed this pair before the subsequent analysis.
Pneumothorax is present in 3,196 images and a chest tube in 1,423, of which 1,253 also show pneumothorax.
By the definitions of Section~\ref{sec:defs}, these counts give $\pi_1 \approx 0.88$, $\pi_0 \approx 0.11$, and $\beta \approx 0.77$, a strong but imperfect association.

\paragraph{CelebA}
CelebA contains 202,599 face images of 10,177 identities, each annotated with 40 binary attributes~\cite{liu2015celeba}.
Any attribute can serve as a proxy sensitive attribute $Z$ for a task defined by another attribute $W$, as in work on spurious correlations and attribute erasure~\cite{sagawa2020groupdro,eid2026leapsc}.
We split identities into a development set of 20\% (2,035 identities), which we used for all exploratory analyses and for forming our hypotheses, and a confirmation set of 80\% (8,142 identities, 161,840 images), which remained sealed until the main experiment.
We use the 34 attributes whose positive rate in the development set is at least 5\% and a fixed set of 100 directed pairs $(W, Z)$ chosen before the main experiment.
A pair enters the analysis only if every test fold of the confirmation set (Section~\ref{sec:splits}) contains at least 100 images in each of the four cells of $W \times Z$, which leaves 85 pairs with $|\beta|$ between $0$ and $0.684$. 


\paragraph{Foundation models}
All encoders are frozen.
For CANDID-PTX, the primary model is RAD-DINO, a self-supervised encoder trained on chest radiographs, and the secondary models are the visual encoder of BiomedCLIP, a biomedical language--image model~\cite{zhang2025biomedclip}, and the general-purpose DINOv2-base~\cite{oquab2024dinov2}.
For CelebA, we use four general-purpose encoders that cover the main pretraining objectives, namely language--image contrastive learning with CLIP ViT-B/32~\cite{radford2021clip} and SigLIP-base~\cite{zhai2023siglip}, self-distillation with DINOv2-base, and masked reconstruction with MAE-base~\cite{he2022mae}.
For RAD-DINO, DINOv2, MAE, CLIP, and BiomedCLIP, we take the CLS token after the final layer normalization, with patch masking disabled for MAE and before the projection into the joint image--text space for CLIP and BiomedCLIP, as in the linear-probe protocols of CLIP~\cite{radford2021clip} and MAE~\cite{he2022mae}.
SigLIP's image tower has no separate projection layer and ends in a multi-head attention-pooling head, whose output we use.
All representations have 768 dimensions and are not $\ell_2$-normalized, and all eraser and covariance computations use double precision.

\begin{table}[t]
\centering
\caption{Data roles within each fold. In the erased conditions, every model other than the eraser sees only erased embeddings.}
\label{tab:roles}
\small
\begin{tabular}{@{}lccc@{}}
\toprule
Model & Train & Select & Evaluate \\
\midrule
Eraser & $A_{\mathrm{train}}$ & -- & -- \\
In-sample probe for $Z$ & $A_{\mathrm{train}}$ & -- & $A_{\mathrm{train}}$ \\
Utility probes for $W$ & $A_{\mathrm{train}}$ & $A_{\mathrm{val}}$ & test fold \\
Attackers for $Z$ & $B_{\mathrm{train}}$ & $B_{\mathrm{val}}$ & test fold \\
\bottomrule
\end{tabular}
\end{table}

\subsection{Data Roles and Splits}
\label{sec:splits}
We use five-fold cross-validation with folds grouped by identity (CelebA) or by patient (CANDID-PTX).
Within the training part of each fold, we split identities or patients into a part $A$ (80\%) and a part $B$ (20\%), and we hold out 10\% of each as validation data, which yields $A_{\mathrm{train}}$, $A_{\mathrm{val}}$, $B_{\mathrm{train}}$, and $B_{\mathrm{val}}$.
Table~\ref{tab:roles} lists how each part is used.
The eraser never sees the data of the attacker, and no model sees the test fold during training or selection.
We compare three conditions: raw embeddings, LEACE as the global eraser of Section~\ref{sec:bg-erasure}, and \textsc{TC-LEACE}.
Section~\ref{sec:case} also evaluates SPLINCE and post-hoc orthogonalization~\cite{weber2025orthogonalization} in analyses outside the fixed plan.

We require at least 50 samples in each group of $Z$ within each stratum of $W$ in $A_{\mathrm{train}}$, at least 30 in each class of $Z$ in $B_{\mathrm{train}}$, and at least 20 in each class of the predicted label in the validation sets.
On CANDID-PTX, we compute the conditional AUROC only when each stratum of the test fold contains at least 20 samples per class of $Z$.

\subsection{Attackers and Utility Probes}
\label{sec:attackers}
Following Section~\ref{sec:threat}, the unconditional attacker receives $E'$, and the conditional attacker is a single model that also receives $W$ as a binary input, standardized with $E'$ on $B_{\mathrm{train}}$.
A linear conditional attacker thus applies one direction in $E'$ to both strata, whereas the MLP can also model interactions between $W$ and $E'$.
For each attacker type, we train logistic regression and a fixed-configuration MLP averaged over three seeds (Appendix~\ref{app:details}), and the \emph{combined attacker} selects the better one in each fold on $B_{\mathrm{val}}$.
The conditional combined attacker selects by conditional AUROC, or by unconditional AUROC when validation strata are too small, as on CANDID-PTX (Appendix~\ref{app:plan}).
Unless noted, we report the combined attackers.
A logistic-regression probe for $Z$, trained and evaluated on the erased fitting data, reproduces the probe-to-chance audit (Section~\ref{sec:audits}).
Recipients are modeled by linear and MLP probes for $W$.

\subsection{Metrics}
\label{sec:metrics}
We compute every AUROC on each fold's evaluation data (Table~\ref{tab:roles}), scoring ties as $0.5$, and average across folds weighted by the number of positive--negative pairs, without pooling scores.
We report the unconditional AUROC, the conditional AUROC of Eq.~\eqref{eq:condauc}, and the difference between the unconditional AUROC of the unconditional attacker and the conditional AUROC of the conditional attacker.
This difference is positive when a population-level audit reports more inference than the conditional attacker achieves within strata.
Because the two attackers differ, Eq.~\eqref{eq:gap} links this difference to imputation only approximately.
On CelebA, we summarize its dependence on $|\beta|$ by the least-squares slope across pairs.
Retention of conditional leakage under an eraser is $(\mathrm{AUC}_{\mathrm{cond}}^{\mathrm{erased}} - 0.5)/(\mathrm{AUC}_{\mathrm{cond}}^{\mathrm{raw}} - 0.5)$.
A probe's utility loss is the drop in its task AUROC relative to raw embeddings, which we split within each fold into within-group and cross-group parts with Eq.~\eqref{eq:decomp} ($Y=W$, $G=Z$), so that both parts sum to the total loss.
We evaluate Corollary~\ref{cor:xauc} in two forms, a guardedness prediction that applies Eq.~\eqref{eq:xauc} to the within-group AUROCs after erasure and an ex-ante prediction that applies it to those before erasure, both with $\varepsilon = 0$.
Eq.~\eqref{eq:decomp} turns either into a predicted cross-group loss, scored by mean absolute error and $R^2$ across CelebA pairs.

\subsection{Hypotheses and Statistical Analysis}
\label{sec:stats}
Before the main experiment, we fixed the hypotheses for each research question and the criteria under which they count as supported, namely a boundary for each hypothesis and the statistical decision rule (Appendix~\ref{app:plan}).

\noindent\textbf{RQ1:} Do population-level audits miss conditional leakage?
\textbf{H1} (both datasets): after LEACE, the conditional AUROC exceeds $0.5 + \delta_T$, where the margin $\delta_T$ is the noise level of the attack under a permutation null (Appendix~\ref{app:plan}).

\noindent\textbf{RQ2:} Does unconditional inference count imputation as leakage?
\textbf{H2} (CelebA): after LEACE, the difference between unconditional and conditional AUROC has a positive slope against $|\beta|$.
\textbf{H2$'$} (CANDID-PTX): this difference is positive.

\noindent\textbf{RQ3:} Where does the cost fall, and can it be predicted?
\textbf{H3} (CelebA, descriptive): for the linear probe after LEACE, the guardedness prediction of the cross-group loss is more accurate than both a zero prediction and a leave-one-pair-out regression on $|\beta|$.

\noindent\textbf{RQ4:} Does changing the target reduce the excess cost without increasing conditional leakage by more than $\delta_T$?
\textbf{H4} (both datasets): compared with LEACE, \textsc{TC-LEACE} has a lower cross-group loss for the linear probe (H4a) and a conditional AUROC less than $\delta_T$ higher (H4b). 
For H4b, $\delta_T$ is a detection margin, not a tolerance for additional leakage.

For each test, we compute a centered bootstrap $p$-value against its boundary from 1,000 resamples of identities or patients per test fold.
A hypothesis is supported only if its Holm-adjusted $p$-value is below $0.05$, with Holm families of 12 tests on CelebA (H1, H2, and H4 for four models) and 3 tests on CANDID-PTX (H1, H2$'$, and H4 for the primary model RAD-DINO), and all other quantities are descriptive.
On CelebA, H1 and H4 concern averages over pairs, and each resample also redraws the attributes that serve as $Z$.
\begin{table}[t]
\centering
\small
\setlength{\tabcolsep}{4pt}
\begin{threeparttable}
\caption{Results on CANDID-PTX with RAD-DINO embeddings ($W$: pneumothorax, $Z$: chest tube).}
\label{tab:candid}
\begin{tabular}{@{}lccc@{}}
\toprule
 & Raw & LEACE & \textsc{TC-LEACE} \\
\midrule
\multicolumn{4}{@{}l}{\emph{Population-level inference (includes imputation)}} \\
Unconditional AUROC & 0.934 & 0.804 & 0.857 \\
 & {\scriptsize[0.922, 0.944]} & {\scriptsize[0.782, 0.824]} & {\scriptsize[0.839, 0.873]} \\
Uncond.\ $-$ cond. & 0.064 & 0.033 & 0.091 \\
 & {\scriptsize[0.050, 0.078]} & {\scriptsize[0.013, 0.054]} & {\scriptsize[0.072, 0.109]} \\
\midrule
\multicolumn{4}{@{}l}{\emph{Conditional leakage (excludes imputation)}} \\
Conditional AUROC $\downarrow$ & 0.870 & 0.771 & 0.767 \\
 & {\scriptsize[0.850, 0.889]} & {\scriptsize[0.744, 0.797]} & {\scriptsize[0.742, 0.791]} \\
\quad logistic attacker $\downarrow$ & 0.822 & \textbf{0.606} & 0.624 \\
 & {\scriptsize[0.796, 0.846]} & {\scriptsize[0.572, 0.634]} & {\scriptsize[0.593, 0.652]} \\
\midrule
\multicolumn{4}{@{}l}{\emph{Authorized task}} \\
Linear probe AUROC $\uparrow$ & 0.907 & 0.731 & \textbf{0.865} \\
MLP probe AUROC $\uparrow$ & 0.918 & 0.897 & \textbf{0.903} \\
$\mathrm{AUC}_{00}$, linear $\uparrow$ & 0.886 & 0.808 & \textbf{0.854} \\
$\mathrm{AUC}_{10} - \mathrm{AUC}_{00}$, linear & 0.063 & $-0.194$ & 0.033 \\
Predicted imputation & 0.909 & 0.569 & 0.835 \\
\midrule
\multicolumn{4}{@{}l}{\emph{Utility loss relative to raw}} \\
Linear probe $\downarrow$ & 0 & 0.176 & \textbf{0.042} \\
 & & {\scriptsize[0.159, 0.192]} & {\scriptsize[0.037, 0.048]} \\
\quad cross-group part $\downarrow$ & 0 & 0.128 & \textbf{0.023} \\
 & & {\scriptsize[0.115, 0.144]} & {\scriptsize[0.019, 0.027]} \\
MLP probe $\downarrow$ & 0 & 0.021 & \textbf{0.015} \\
 & & {\scriptsize[0.016, 0.027]} & {\scriptsize[0.011, 0.019]} \\
\bottomrule
\end{tabular}
\begin{tablenotes}[flushleft]
\footnotesize
\item Attack AUROCs use the combined attacker unless noted.
Brackets give 95\% bootstrap intervals.
Arrows mark the preferred direction, and rows without arrows are diagnostic.
Bold marks the better of the two erasers where the 95\% interval of their difference excludes zero.
Population-level inference includes the imputation that the task label permits, so it does not isolate leakage.
The cross-group part follows Eq.~\eqref{eq:decomp} with $Y=W$ and $G=Z$.
$\mathrm{AUC}_{00}$ compares untreated pneumothoraces with images without pneumothorax or tube, and $\mathrm{AUC}_{10}$ uses treated pneumothoraces instead.
Predicted imputation is the AUROC of the linear probe's task score as a score for $Z$.
The imputation ceiling is $0.886$ $[0.875, 0.896]$.
\end{tablenotes}
\end{threeparttable}
\end{table}

\section{A Realistic Release: Treatment History in Chest Radiographs}
\label{sec:case}

Each fold contains 3,174 images without pneumothorax or tube, 34 with a tube but no pneumothorax, 389 with pneumothorax but no tube, and 250 to 252 with both.
All five folds meet eligibility rules, and all 1,000 bootstrap resamples are valid.
All three hypotheses are supported, with Holm-adjusted $p \le 0.003$, the bootstrap resolution limit.
Unlike on CelebA, these tests use data our exploratory runs had seen, and those motivated H2$'$ (Appendix~\ref{app:plan}).


\paragraph{RQ1: Population-level audits miss conditional leakage}
After LEACE, both audits of Section~\ref{sec:audits} pass.
A linear probe trained and evaluated on the fitting data attains exactly $0.5$ in every fold and encoder, as LEACE guarantees (Section~\ref{sec:bg-erasure}).
The unconditional attack reaches $0.804$, below the $0.886$ imputation ceiling, and even the recipient's MLP task score, ranking tube status at $0.880$, does not exceed it.
The label-only comparison therefore detects no leakage beyond the diagnosis, but an attack below the ceiling cannot rule such leakage out (Section~\ref{sec:audits}).
The conditional attacker separates tube placement within pneumothorax strata, with a conditional AUROC of $0.771$ whose interval $[0.744, 0.797]$ lies above $0.5 + \delta_T = 0.578$, so H1 is supported.
Both audits thus pass a release that still discloses treatment history within strata.
Even the logistic-regression attacker reaches $0.606$ within strata, as Eq.~\eqref{eq:guarddecomp} predicts for a release that keeps its linear task signal, and the first post-hoc analysis separates this residual from finite-sample error.

\paragraph{RQ2: Unconditional inference counts imputation as leakage}
The unconditional AUROC exceeds the conditional AUROC by $0.033$ $[0.013, 0.054]$ after LEACE, so H2$'$ is supported, and by $0.064$ $[0.050, 0.078]$ on raw embeddings.
This gap is consistent with an unconditional audit counting part of the permitted inference as leakage, and its size depends on the eraser.
After \textsc{TC-LEACE}, the difference reaches $0.091$ $[0.072, 0.109]$, because \textsc{TC-LEACE} keeps the permitted association (Eq.~\eqref{eq:tcglobal}).
Such an audit would therefore wrongly rate \textsc{TC-LEACE} as the leakier eraser, although their conditional AUROCs nearly match.

\paragraph{RQ3: The cost falls on linear readers and cross-group pairs}
LEACE reduces the linear pneumothorax probe's AUROC by $0.176$, and $73\%$ of this loss falls on cross-group pairs.
The guardedness and ex-ante predictions of Corollary~\ref{cor:xauc} put this part at $0.153$ and $0.146$ and overestimate it, because guardedness holds only partly on new data.
Adding the measured residual $\varepsilon$ reduces the guardedness prediction's error to $0.002$.
The shift that Proposition~\ref{prop:means} forces is visible in the task score.
On raw embeddings, the linear score ranks tube status above the ceiling, consistent with chest tubes acting as a shortcut~\cite{oakdenrayner2020hidden}.
After LEACE, it ranks tube status at only $0.569$, and treated pneumothoraces become harder to detect than untreated ones, as in Figure~\ref{fig:concept}(b).
The MLP probe, in contrast, loses only $0.021$.
LEACE thus hides the permitted association from linear readers rather than removing it, and the cost falls mainly on linear recipients, the only readers that LEACE's guarantee covers.

\paragraph{RQ4: Changing the target removes most of the cost}
\textsc{TC-LEACE} keeps the conditional AUROC at the level of LEACE (difference $-0.004$ $[-0.021, 0.014]$, well below $\delta_T = 0.0785$) but cuts the linear utility loss by $76\%$, from $0.176$ to $0.042$, and H4 is supported (difference in cross-group loss $-0.105$ $[-0.118, -0.093]$).
Only about half of its remaining linear loss falls on cross-group pairs ($0.023$ of $0.042$), and $\mathrm{AUC}_{10} - \mathrm{AUC}_{00}$ stays positive ($0.033$), so the forced shift of global guardedness does not arise.
The MLP probe's loss also falls, from $0.021$ to $0.015$.
The gain does not come only from preserving the shortcut, since \textsc{TC-LEACE} also outperforms LEACE on pairs without any chest tube ($\mathrm{AUC}_{00}$, difference $0.046$ $[0.038, 0.055]$).
The additional utility that LEACE sacrifices thus buys protection against permitted inference, not a reduction of conditional leakage.
Neither eraser stops nonlinear attackers.
The logistic-regression attacker is even slightly stronger after \textsc{TC-LEACE} (difference $0.018$ $[0.004, 0.033]$), but only in the stratum without pneumothorax ($0.651$ versus $0.608$), whose erasure direction rests on about 100 positives in 768 dimensions, as the first post-hoc analysis indicates.

\paragraph{Secondary models}
DINOv2 and BiomedCLIP show the same pattern (Table~\ref{tab:candid-secondary}).
After LEACE, their conditional AUROCs remain at $0.671$ and $0.648$, so conditional leakage persists in every encoder.
The combined attacker understates this leakage, since its fallback selection (Section~\ref{sec:attackers}) picks logistic regression in some folds, and the MLP attacker alone reaches $0.723$ and $0.731$.
\textsc{TC-LEACE} cuts their linear utility loss by about three quarters.
Its conditional AUROC is $0.025$ $[0.003, 0.046]$ and $0.023$ $[0.007, 0.038]$ higher than that of LEACE, a detectable increase that stays below $\delta_T$, whereas the MLP attacker alone differs by $0.010$ and $0.004$, with intervals including zero.

\paragraph{Post-hoc analyses}
After the main experiment, we ran three exploratory analyses on CANDID-PTX (Table~\ref{tab:posthoc}).
First, when the custodian fits the eraser on the released embeddings, LEACE's global guarantee holds on them, yet a logistic-regression attacker given $W$ still reaches $0.564$ and $0.530$ within the two strata for RAD-DINO, whereas \textsc{TC-LEACE} lowers them to $0.502$ and $0.483$.
This isolates the residual of Lemma~\ref{lem:residual} from finite-sample error.
Second, in both fitting settings, SPLINCE differs from LEACE by at most $0.009$ for retrained linear readers, as their shared kernel implies (Section~\ref{sec:readers}).
Third, the orthogonalization proposed for medical embeddings~\cite{weber2025orthogonalization} subtracts the global mean difference, including the $W$-mediated component of Eq.~\eqref{eq:covdecomp}, and raises the MLP attacker's conditional AUROC by $0.066$ to $0.142$ over raw embeddings across encoders.
A defense designed to remove sensitive information thus increases its disclosure within task strata under strong association.
\section{Generality Across Attributes and Models}
\label{sec:scale}

We now ask whether the effects of Section~\ref{sec:case} generalize and how they depend on the association $|\beta|$.
Of the 100 directed pairs $(W, Z)$ that we fixed over 34 CelebA attributes before the main experiment, 85 meet the eligibility rule of Section~\ref{sec:data}, and we evaluate each with four encoders.
For descriptive analysis, we split these pairs at $|\beta| = 0.0256$ and $0.11$ into a low, a middle, and a high tertile of 29, 28, and 28 pairs, with mean $|\beta|$ of $0.013$, $0.063$, and $0.226$.

\begin{table}[t]
\centering
\small
\setlength{\tabcolsep}{3pt}
\begin{threeparttable}
\caption{Results on CelebA after LEACE.}
\label{tab:celeba}
\begin{tabular}{@{}lcccc@{}}
\toprule
Model & Cond.\ AUROC & Retention & H2 slope & Diff. \\
\midrule
CLIP & 0.900 & 96.3\% & 0.064 & 0.0029 \\
 & {\scriptsize[0.863, 0.933]} & {\scriptsize[94.9, 97.4]} & {\scriptsize[0.017, 0.107]} & {\scriptsize[0.0010, 0.0055]} \\
DINOv2 & 0.895 & 97.3\% & 0.065 & 0.0029 \\
 & {\scriptsize[0.858, 0.926]} & {\scriptsize[96.6, 97.8]} & {\scriptsize[0.018, 0.107]} & {\scriptsize[0.0011, 0.0055]} \\
SigLIP & 0.899 & 95.5\% & 0.068 & 0.0028 \\
 & {\scriptsize[0.862, 0.931]} & {\scriptsize[93.9, 96.8]} & {\scriptsize[0.018, 0.116]} & {\scriptsize[0.0010, 0.0055]} \\
MAE & 0.860 & 91.5\% & 0.077 & 0.0034 \\
 & {\scriptsize[0.820, 0.893]} & {\scriptsize[90.1, 92.8]} & {\scriptsize[0.025, 0.120]} & {\scriptsize[0.0013, 0.0061]} \\
\bottomrule
\end{tabular}
\begin{tablenotes}[flushleft]
\footnotesize
\item Attack AUROCs use the combined attacker.
Cond.\ AUROC is the conditional AUROC averaged over pairs (H1).
Retention is the share of the conditional leakage of raw embeddings that LEACE retains (Section~\ref{sec:metrics}).
H2 slope is the slope of the unconditional minus conditional AUROC against $|\beta|$, and Diff.\ is the mean of this difference over pairs.
Brackets give 95\% bootstrap intervals.
\end{tablenotes}
\end{threeparttable}
\end{table}

\begin{figure}[htbp]
\centering
\includegraphics[width=\columnwidth]{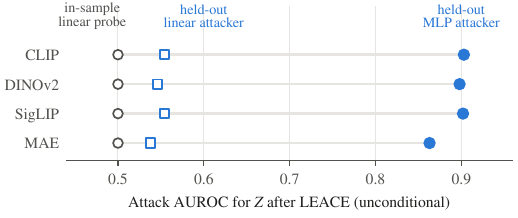}
\caption{Three views of the same CelebA embeddings after LEACE, averaged over the 85 pairs. The in-sample linear probe sits at $0.5$ by construction, the held-out linear attacker stays slightly above chance, and the held-out MLP attacker still recovers the attribute.}
\label{fig:audit}
\end{figure}
\paragraph{RQ1: Conditional leakage persists in every encoder}
Figure~\ref{fig:audit} contrasts three views of the same erased embeddings.
A linear probe trained and evaluated on the fitting data attains an AUROC of exactly $0.5$ in every pair, fold, and encoder, so the probe-to-chance audit passes.
On new data, the unconditional linear attacker reaches only $0.538$--$0.554$, whereas the unconditional MLP attacker reaches $0.863$--$0.903$.
After LEACE, the conditional AUROC lies between $0.860$ and $0.900$ (Table~\ref{tab:celeba}), with lower interval bounds of at least $0.820$, far above $0.5 + \delta_T = 0.523$, and H1 is supported.
LEACE retains between $91.5\%$ and $97.3\%$ of the conditional leakage of raw embeddings.
The linear part grows with the association, as Eq.~\eqref{eq:guarddecomp} predicts.
For CLIP, a logistic-regression attacker that knows $W$ reaches $0.579$ $[0.568, 0.590]$ after LEACE and $0.619$ in the high tertile, and the other models behave alike.

\newpage
\begin{figure}[htbp]
\centering
\includegraphics[width=\columnwidth]{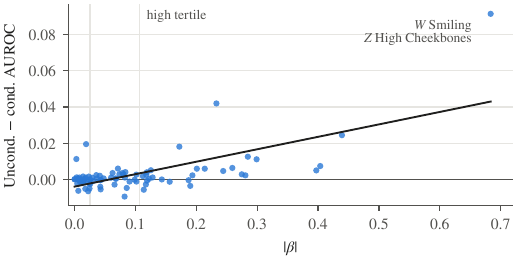}
\caption{Difference between unconditional and conditional AUROC after LEACE for each of the 85 CelebA pairs, averaged over the four encoders. The line is a least-squares fit whose slope equals the mean of the four slopes that H2 tests. Vertical lines mark the tertile boundaries of $|\beta|$.}
\label{fig:imputation}
\end{figure}

\paragraph{RQ2: Unconditional inference counts more imputation as $|\beta|$ grows}
After LEACE, the difference between unconditional and conditional AUROC grows with $|\beta|$, with slopes between $0.064$ and $0.077$ whose intervals exclude zero (Table~\ref{tab:celeba}), and H2 is supported (Figure~\ref{fig:imputation}).
Without the pair with the strongest association, with Smiling as $W$ and High Cheekbones as $Z$, the slope falls to between $0.030$ and $0.044$ but remains positive.
The effect is small on average, with a mean difference of about $0.003$ over all pairs, $0.008$ to $0.010$ in the high tertile, and nearly zero in the low tertile.
The discrepancy thus matters mainly under strong association, as in CANDID-PTX.

\begin{figure}[htbp]
\centering
\includegraphics[width=\columnwidth]{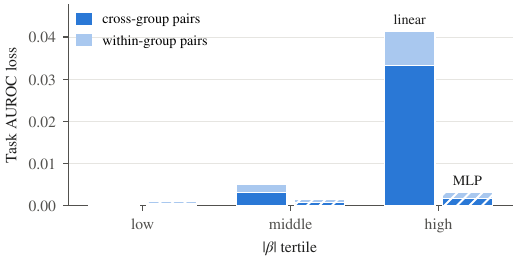}
\caption{Utility loss after LEACE by tertile of $|\beta|$, split into cross-group and within-group pairs, for linear probes (solid) and MLP probes (hatched), averaged over the four encoders. The loss concentrates on cross-group pairs and on linear readers.}
\label{fig:cost}
\end{figure}

\paragraph{RQ3: The cost falls on linear readers and is predictable}
After LEACE, the cross-group loss is $0.011$ to $0.014$ for the linear probe and $0.0006$ to $0.0017$ for the MLP probe over all pairs, and $0.031$ to $0.039$ and $0.0010$ to $0.0031$ in the high tertile.
For the linear probe, this part accounts for $76\%$ to $80\%$ of the loss overall, close to the $73\%$ on CANDID-PTX, and its share rises from under one half in the low tertile to about four fifths in the high tertile (Figure~\ref{fig:cost}), consistent with Section~\ref{sec:fail1}.
The guardedness prediction of Corollary~\ref{cor:xauc} errs by $0.0018$ to $0.0029$ on average ($R^2$ of $0.97$ to $0.99$), whereas predicting zero errs by the loss itself and a leave-one-pair-out regression on $|\beta|$ by $0.014$ ($R^2$ of $0.41$ to $0.53$).
Both improvements have intervals that exclude zero, so H3 holds for all four encoders.
The ex-ante prediction, which uses only within-group AUROCs before erasure, comes close, with errors of $0.0027$ to $0.0033$ and $R^2$ of $0.92$ to $0.96$, so a custodian can estimate the cost in advance.
The corollary strongly overestimates the loss of MLP probes and of \textsc{TC-LEACE}, since neither is subject to the global equal-means constraint.

\begin{figure}[htbp]
\centering
\includegraphics[width=\columnwidth]{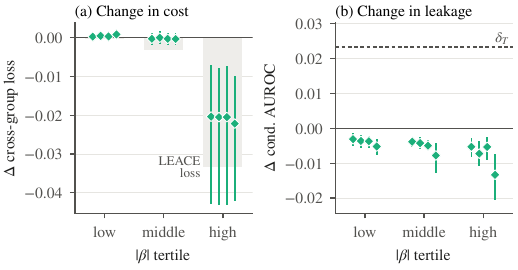}
\caption{Differences $\Delta$ (\textsc{TC-LEACE} minus LEACE) by tertile of $|\beta|$, averaged over the pairs of each tertile, in the cross-group loss of the linear probe for H4a (a) and in conditional AUROC for H4b (b). Negative values favor \textsc{TC-LEACE}. Each marker is one encoder, ordered CLIP, DINOv2, SigLIP, and MAE from left to right, with its 95\% interval, and intervals in (a) use only the valid resamples (Appendix~\ref{app:plan}). Gray bars in (a) show the negative of the cross-group loss under LEACE, averaged over the four encoders, which $\Delta$ would reach if \textsc{TC-LEACE} removed this loss entirely. The dashed line in (b) marks the margin $\delta_T$ of H4b. Both parts are tested over all pairs, and the tertile intervals are descriptive.}
\label{fig:cond}
\end{figure}

\paragraph{RQ4: Changing the target reduces the cost where the association is strong}
\textsc{TC-LEACE} lowers the cross-group loss of the linear probe from $0.011$--$0.014$ to $0.0044$--$0.0069$, with paired differences of $-0.0065$ to $-0.0071$ whose 95\% intervals exclude zero, and the effect concentrates where the theory places it (Figure~\ref{fig:cond}a).
In the high tertile, the difference is $-0.020$ to $-0.022$, about two thirds of the loss.
In the middle tertile it is close to zero, and in the low tertile it is $+0.0003$ to $+0.0009$, with intervals that exclude zero.
This small excess is consistent with \textsc{TC-LEACE} removing up to two directions where LEACE removes one, while the $W$-mediated component that it preserves vanishes as $\beta \to 0$.
\textsc{TC-LEACE}'s advantage thus appears only under strong association, growing with it, in line with its much larger reduction on CANDID-PTX (Section~\ref{sec:case}).
Over all pairs, however, H4a fails, since its $p$-values of $0.014$ to $0.039$ become $0.056$ to $0.109$ after the Holm correction over twelve tests, whereas the intervals above are unadjusted.
The conditional AUROC of \textsc{TC-LEACE} is $0.004$ to $0.009$ lower than that of LEACE over all pairs, and lower in every tertile (Figure~\ref{fig:cond}b), so H4b holds by a wide margin.

\paragraph{Summary}
Table~\ref{tab:findings} collects the answers on both datasets.
For RQ1, the probe-to-chance audit misses conditional leakage on both datasets and the label-only comparison misses it on CANDID-PTX, while it persists after LEACE in every encoder (H1).
For RQ2, unconditional evaluation counts permitted imputation as leakage, substantially on CANDID-PTX but little on the mostly weakly associated CelebA pairs (H2, H2$'$).
For RQ3, the cost of global erasure falls on linear readers and cross-group pairs in both settings, and on CelebA the guardedness and ex-ante predictions track it (H3).
For RQ4, changing the target removes most of this cost on CANDID-PTX (H4).
On CelebA, H4 is not supported, since H4a fails when tested over the full set of pairs, most of which are weakly associated, while H4b holds, although both parts hold descriptively for strongly associated pairs.

\begin{table}[t]
\centering
\footnotesize
\setlength{\tabcolsep}{3pt}
\caption{Findings by research question on both datasets.}
\label{tab:findings}
\begin{tabular}{@{}L{0.08\columnwidth}L{0.43\columnwidth}L{0.43\columnwidth}@{}}
\toprule
 & CANDID-PTX (RAD-DINO) & CelebA (85 pairs, 4 encoders) \\
\midrule
RQ1 & Both audits pass, yet the conditional AUROC is $0.771$ after LEACE. \textbf{H1 supported.} & Conditional AUROC $0.860$--$0.900$ after LEACE. \textbf{H1 supported} for all encoders. \\
\addlinespace
RQ2 & Unconditional minus conditional AUROC $0.033$ after LEACE and $0.091$ after \textsc{TC-LEACE}. \textbf{H2$'$ supported.} & The difference grows with $|\beta|$ but averages only $0.003$. \textbf{H2 supported.} \\
\addlinespace
RQ3 & $73\%$ of the linear loss falls on cross-group pairs, and Corollary~\ref{cor:xauc} overestimates this part. \textbf{H3 not tested} (one pair). & The guardedness prediction errs by $0.0018$--$0.0029$, and the MLP probe loses little. \textbf{H3 holds} (multiple pairs). \\
\addlinespace
RQ4 & Linear loss falls from 0.176 to 0.042, with conditional AUROC within $\delta_T$ of LEACE's. \textbf{H4 supported}. & The cross-group loss falls mainly in the high tertile. \textbf{H4a fails, and H4b holds}. \\
\bottomrule
\end{tabular}
\end{table}
\section{Discussion}
\label{sec:discussion}

\subsection{Recommendations for Data Custodians}
\label{sec:recommend}

Our results suggest four practices for releasing embeddings under an authorized task.
\begin{enumerate}
\item \textbf{Audit conditionally.}
Report the conditional AUROC of nonlinear attackers, the AUROC of each stratum, and the imputation ceiling, since neither a probe at chance on the fitting data nor an attack below a label-only baseline shows that the release discloses nothing beyond the task.
\item \textbf{Estimate the cost before erasing.}
The ex-ante prediction of Corollary~\ref{cor:xauc} estimates the cost of global erasure for linear recipients from raw embeddings, and it is conservative where guardedness holds only partly on new data, as on CANDID-PTX.
\item \textbf{Match the target to the authorization.}
Under a strongly associated authorized task, erase within the strata of the task rather than globally, avoid residualizing on the global mean difference, and ensure enough samples in every cell of $W \times Z$, since the eraser estimates one direction per stratum.
\item \textbf{Treat linear guarantees as linear.}
Neither LEACE nor \textsc{TC-LEACE} constrains nonlinear readers, so if nonlinear attackers are in scope, the custodian needs nonlinear erasure with a conditional target or controls outside the representation, such as access restrictions and data use agreements.
\end{enumerate}

\newpage
\subsection{Limitations}
\label{sec:limits}

The practical significance of our findings lies mainly in strongly associated pairs and in the medical case.
Our evidence from a realistic release rests on a single prespecified attribute pair in CANDID-PTX, whose stratum without pneumothorax contains few images with a tube, 34 per test fold and 1 to 9 in the attacker's validation data, so the conditional attacker had to be selected by unconditional performance.
The CelebA attributes are not sensitive in themselves and serve only as proxies.
We study binary task labels and attributes, and the closed-form prediction relies on a binormal approximation with equal variances.
We evaluate linear erasers only and do not provide a nonlinear eraser with a conditional target.
Our bootstrap reflects uncertainty in the test data and, on CelebA, in the choice of attributes, but not variation from retraining the probes and attackers.
\textsc{TC-LEACE} guarantees linear guardedness within each stratum of its fitting data, but the rare stratum of CANDID-PTX gives it only about 100 positives to estimate its direction, so the guarantee holds only approximately for embeddings outside the fitting data.
On CelebA, the bootstrap re-applied the cell-size rule of the pair selection to every resample, which left only 356 of 1,000 resamples valid for H4a.
Valid and invalid resamples show nearly identical distributions for H1 and H4b and differ by about 0.15 standard deviations for H2, so the resulting selection bias appears small.

\section{Related Work}
\label{sec:related}

\paragraph{Privacy of shared representations}
Embeddings released for one purpose can disclose other information through attribute inference, inversion, and feature leakage~\cite{song2020embedding,morris2023embeddings,melis2019exploiting}.
In medical imaging, deep models recognize race from radiographs~\cite{gichoya2022race}, chest radiograph foundation models encode protected characteristics~\cite{glocker2023risk}, FM embeddings retain demographic attributes~\cite{santos2025persistent,kim2026encoding}, and brain MRI embeddings enable linkage and attribute inference~\cite{bajar2026brainmri}.
Post-hoc orthogonalization has been proposed to remove protected features from chest radiograph embeddings~\cite{weber2025orthogonalization}, but it can increase conditional leakage under strong association (Section~\ref{sec:case}).

\paragraph{What counts as leakage}
Contextual integrity holds that whether an information flow violates privacy depends on the context and purpose of the flow~\cite{nissenbaum2004privacy}.
Differential privacy likewise permits inferences about an individual that follow from population-level statistics rather than from that individual's data~\cite{dwork2014algorithmic}.
In the same spirit, Jayaraman and Evans argue that attribute inference should be measured beyond imputation~\cite{jayaraman2022imputation}, and Stadler et al.\ observe that useful representations must reveal what the task label reveals~\cite{stadler2024leastprivilege}.
They also prove that limiting the leakage of every other attribute to this level conflicts with high utility, whereas we target a single specified attribute and turn the distinction into an audit of erased embeddings.
Existing baselines compare attribute inference with what ground-truth labels reveal~\cite{wang2019balanced,eid2026leapsc}, but they remain unconditional and can miss conditional leakage (Section~\ref{sec:case}).
Our conditional AUROC instead follows covariate-adjusted ROC analysis, which separates comparisons within covariate strata from pooled ones~\cite{janes2009covariate}.

\paragraph{Concept erasure}
Beyond the erasers of Section~\ref{sec:bg-erasure}, linear erasers include mean projection~\cite{haghighatkhah2022better}, SAL~\cite{shao2023sal}, and Fair PCA~\cite{kleindessner2023fairpca}.
Adversarial removal~\cite{coavoux2018privacy,elazar2018adversarial} and the debiasing of word embeddings~\cite{gonen2019lipstick} can leave the attribute recoverable, kernel erasure does not transfer across adversaries~\cite{ravfogel2022kernel}, and nonlinear erasers pursue $E' \perp Z$~\cite{akbari2026obliviator,saillenfest2025density,simai2026principled}, whose limits are characterized~\cite{chowdhury2025fundamental}.
All of these methods target a global constraint that does not condition on the authorized task label.
The closest work fits one LEACE projector per task class within an encoder trained for semantic communication and selects it with a predicted label~\cite{eid2026leapsc}.
In contrast, \textsc{TC-LEACE} applies a single map post hoc to frozen embeddings and needs no labels at release, and our analysis explains why the global target costs utility.

\paragraph{Fair representations and conditional independence}
Fair representations remove a protected attribute while keeping task information~\cite{zemel2013lfr,edwards2016censoring}, but parity-style constraints incur an unavoidable error when base rates differ across groups ($\beta \neq 0$)~\cite{hardt2016equality,zhao2019inherent}.
Global and task-conditional guardedness are the linear analogues of demographic parity and equalized odds, and Section~\ref{sec:mechanism} locates where the cost of the parity-style target falls.
Conditioning on the label is the standard remedy, through adversarial debiasing~\cite{zhang2018adversarial}, equalized-odds objectives~\cite{madras2018laftr}, conditional fair representations~\cite{zhao2020cfair}, and conditional constraints in medical imaging and latent spaces~\cite{adeli2021representation,hwa2024enforcing}.
These methods shape the representation during training, whereas we erase fixed FM embeddings post hoc for privacy under an authorized task.
The analysis suggests that globally erasing race or sex from medical embeddings also suppresses genuine differences in prevalence, with the cost falling on the same cross-group pairs.
\section{Conclusion}
\label{sec:conclusion}

Concept erasure is usually designed and audited at the level of the whole population.
When embeddings are shared for an authorized task that is associated with a sensitive attribute, this target is misaligned with the privacy requirement in two ways.
It spends utility to suppress inference that the task label already permits, and it does not constrain what the embeddings reveal within each task label.
We proved that global and task-conditional linear guardedness are incompatible under such association when a linear task signal is preserved, and we showed that the cost of global guardedness falls on linear readers and cross-group pairs and can be estimated before erasure.
On chest radiographs, population-level audits pass LEACE-erased embeddings that still disclose treatment history among patients with the same diagnosis, and \textsc{TC-LEACE} removes most of the linear utility cost at conditional leakage within a prespecified margin of LEACE's, although neither eraser stops nonlinear attackers.
On CelebA, its gains concentrate in strongly associated pairs.
Our work therefore argues that, when embeddings are released for a specific purpose, audits should be conditional and erasure targets aligned with the authorized task.
\clearpage
\bibliographystyle{IEEEtran}
\bibliography{references}

\clearpage
\appendices

\section{Proofs}
\label{app:proofs}

\paragraph{Imputation ceiling under conditional independence}
Let $s(E',W)$ be any score, possibly randomized, and assume $E' \perp Z \mid W$.
Given $Z=z$, the label $W$ equals $1$ with probability $\pi_z$, and given $W=w$ the score has a law $G_w$ that does not depend on $z$.
For independent draws $S_g \sim G_g$ and $S'_h \sim G_h$, let $A_{gh} = \Pr[S_g > S'_h] + \tfrac12\Pr[S_g = S'_h]$, so that $A_{00} = A_{11} = \tfrac12$ and $A_{01} = 1 - A_{10}$.
Conditioning Eq.~\eqref{eq:auc} with $Y=Z$ on the task labels of the two draws gives
\[
\begin{aligned}
\mathrm{AUC}(s;Z)
&= \tfrac12 + \big(\pi_1(1-\pi_0) - (1-\pi_1)\pi_0\big)\big(A_{10} - \tfrac12\big)\\
&= \tfrac12 + \beta\big(A_{10} - \tfrac12\big).
\end{aligned}
\]
Since $A_{10} \in [0,1]$, $\mathrm{AUC}(s;Z) \le \tfrac12 + \tfrac{|\beta|}{2} = \mathrm{AUC}^*_{\mathrm{imp}}$, with equality for the score $W$ or $1-W$.
Within each stratum, both groups of $Z$ share the law $G_w$, so the conditional AUROC equals $0.5$.

\paragraph{Eq.~\eqref{eq:shift}}
Within group $Z=z$, a fraction $\pi_z$ of the samples has $W=1$, so $\mathbb{E}[s \mid Z=z] = (1-\pi_z)m_z + \pi_z(m_z + d_z) = m_z + \pi_z d_z$.
Linear guardedness makes $\mathbb{E}[s \mid Z=z] = a^\top \mathbb{E}[E' \mid Z=z] + b$ equal for both values of $z$, and equating the two group means gives Eq.~\eqref{eq:shift}.
The same computation with $\mathbb{E}[s \mid Z=1] - \mathbb{E}[s \mid Z=0] = \sigma\varepsilon$ gives the form with $\varepsilon$ in Section~\ref{sec:fail1}.

\paragraph{Proposition~\ref{prop:means}}
For part~(a), suppose that an affine score $s$ satisfies $\mathbb{E}[s \mid W,Z] = \mathbb{E}[s \mid W]$, and write $g_w = \mathbb{E}[s \mid W=w]$.
Then $\mathbb{E}[s \mid Z=z] = g_0 + \pi_z(g_1 - g_0)$, so the two group means differ by $\beta(g_1 - g_0)$.
Linear guardedness requires this difference to vanish, which with $\beta \neq 0$ forces $g_1 = g_0$.
For part~(b), suppose that $s$ is a measurable score of $E'$ with $s \perp Z \mid W$, and let $f$ be any bounded measurable function.
Writing $h_w = \mathbb{E}[f(s) \mid W=w]$, we obtain $\mathbb{E}[f(s) \mid Z=z] = h_0 + \pi_z(h_1 - h_0)$.
Since $E' \perp Z$ implies $s \perp Z$, these expectations agree for both values of $z$, so $\beta(h_1 - h_0) = 0$ and hence $h_1 = h_0$.
As this holds for every bounded $f$, the law of $s$ is the same given $W=0$ and $W=1$, that is, $s \perp W$.

\paragraph{Corollary~\ref{cor:xauc}}
If two independent normal variables have common variance $\sigma^2$ and means that differ by $\mu$, the probability that the first exceeds the second is $\Phi\big(\mu/(\sigma\sqrt{2})\big)$.
Within group $Z=z$, task positives have mean $m_z + d_z$ and task negatives have mean $m_z$, so $\mathrm{AUC}_{zz} = \Phi(\Delta_z/\sqrt{2})$.
Task positives with $Z=1$ and task negatives with $Z=0$ differ in mean by $m_1 + d_1 - m_0 = \sigma(\Delta_1 + \delta)$, which gives $\mathrm{AUC}_{10}$, and task positives with $Z=0$ and task negatives with $Z=1$ differ by $\sigma(\Delta_0 - \delta)$, which gives $\mathrm{AUC}_{01}$.
Dividing Eq.~\eqref{eq:shift} by $\sigma$ gives $\delta = \pi_0\Delta_0 - \pi_1\Delta_1$, which equals $-\beta\Delta$ when $\Delta_0 = \Delta_1 = \Delta$.

\paragraph{Eqs.~\eqref{eq:covdecomp}, \eqref{eq:guarddecomp}, and~\eqref{eq:tcglobal}}
By the law of total covariance, $\mathrm{Cov}(E, Z) = \mathbb{E}[\mathrm{Cov}(E, Z \mid W)] + \mathrm{Cov}(\mathbb{E}[E \mid W], \mathbb{E}[Z \mid W])$.
The first term equals $(1-q)\,c_0 + q\,c_1$.
Since $\mathbb{E}[E \mid W] = \mu_0 + W(\mu_1 - \mu_0)$ and $\mathbb{E}[Z \mid W] = \rho_0 + W(\rho_1 - \rho_0)$, the second term equals $q(1-q)(\mu_1 - \mu_0)(\rho_1 - \rho_0)$.
The identity holds for any representation, so for $E'$ it gives Eq.~\eqref{eq:guarddecomp} when $\mathrm{Cov}(E', Z) = 0$ and Eq.~\eqref{eq:tcglobal} when $c_0' = c_1' = 0$, using $\mathrm{Cov}(E', W) = q(1-q)(\mu_1' - \mu_0')$.

\paragraph{Range facts}
Let $\Pi$ denote the orthogonal projection onto the range of $\Sigma$, so that $\Sigma^{1/2}\Sigma^{+1/2} = \Sigma^{+1/2}\Sigma^{1/2} = \Pi$.
If $\Sigma v = 0$, then $v^\top E$ is almost surely constant, so $v^\top \Sigma_{EZ} = 0$, and since $\Sigma \succeq \Pr[W=w]\,\mathrm{Cov}(E \mid W=w)$, also $v^\top c_w = 0$.
Hence $\Sigma_{EZ}$, $c_0$, and $c_1$ lie in the range of $\Sigma$, and so do the ranges of $P$ and $P_C$.

\paragraph{Lemma~\ref{lem:residual}}
Eq.~\eqref{eq:leace} can be written as $E' = ME + (I - M)\mu$, so $\mathrm{Cov}(E', Z \mid W=w) = M c_w$.
With $u_w = \Sigma^{+1/2} c_w$, the range facts give $\Sigma^{1/2} u_w = c_w$ and hence $M c_w = \Sigma^{1/2}(I - P)\,u_w$.
The vector $(I - P)\,u_w$ lies in the range of $\Sigma$, on which $\Sigma^{1/2}$ is injective, so $M c_w = 0$ if and only if $u_w$ lies in the range of $P$, which is $\operatorname{span}(\Sigma^{+1/2}\Sigma_{EZ})$.
Applying $\Sigma^{1/2}$ shows that this is equivalent to $c_w \in \operatorname{span}(\Sigma_{EZ})$, and multiplying $M c_w$ by $\Sigma^{+1/2}$ gives the whitened residual $(I - P)\,u_w$.

\paragraph{Proposition~\ref{prop:tc}}
Eq.~\eqref{eq:tc} can be written as $E' = M_C E + (I - M_C)\mu$ with $M_C = I - \Sigma^{1/2} P_C \Sigma^{+1/2}$, so $\mathrm{Cov}(E', Z \mid W=w) = c_w - \Sigma^{1/2} P_C u_w$.
Since $u_w$ lies in the range of $P_C$, this equals $c_w - \Sigma^{1/2} u_w = 0$.
For binary $Z$ with $0 < \rho_w < 1$, $\mathrm{Cov}(E', Z \mid W=w) = \rho_w(1-\rho_w)\big(\mathbb{E}[E' \mid W=w, Z=1] - \mathbb{E}[E' \mid W=w, Z=0]\big)$, so the covariance vanishes exactly when the two conditional means coincide.

\section{Finite-Sample Residual}
\label{app:finite}

We give an order-of-magnitude argument in whitened coordinates, in which the within-class covariance is approximately the identity.
Let $\delta$ be the population difference between the group means of $Z$, and let $\hat\delta = \delta + \xi$ be its estimate from a fitting sample of size $n$ with a fraction $p$ of positives.
Each coordinate of $\xi$ has variance approximately $1/(n\,p(1-p))$.
The eraser removes the direction of $\hat\delta$, so the population difference that remains is $(I - \hat P)\delta = -(I - \hat P)\xi$, where $\hat P$ projects onto $\hat\delta$.
When $\|\delta\|$ is large relative to $\|\xi\|$, its squared norm is approximately $(d-1)/(n\,p(1-p))$, independent of $\delta$.
For Gaussian classes, a linear score aligned with this residual would reach an AUROC of $\Phi\big(\sqrt{(d-1)/(2n\,p(1-p))}\big)$.
This value overstates what a trained attacker achieves, because the attacker must estimate the residual direction from its own data, so we use it only as a rough upper reference.
For \textsc{TC-LEACE}, the same argument applies within each stratum, with $n$ and $p$ replaced by the size of the stratum and $\rho_w$, so a stratum with few samples in one group of $Z$, such as the stratum without pneumothorax in CANDID-PTX, leaves a larger residual.

\section{Analysis Plan and Design Issues}
\label{app:plan}

\paragraph{Plan}
Before the main experiment, we froze an analysis plan that fixes the data splits, the eligibility rules, the attackers and probes, the metrics, the hypotheses H1 to H4 with their boundaries, the decision rule, the margin $\delta_T$, and the Holm families.
The repository contains the plan.
H4 is supported only if both H4a and H4b hold.
Following the plan, its $p$-value is the larger of the two component $p$-values, an intersection--union test, and this joint $p$-value enters the Holm family.

\paragraph{Margin}
The margin is $\delta_T = \max(0.02,\ q_{0.975} - 0.5)$, where $q_{0.975}$ is the 97.5th percentile of the conditional AUROC after LEACE when $Z$ is shuffled within the strata of $W$ in the attacker's training data $B$.
For CelebA, we computed it on the development set with 10 pairs, four encoders, and 10 permutations each, which gives $\delta_T = 0.0233$.
For CANDID-PTX, we computed it with RAD-DINO and 20 permutations after freezing the plan and before examining any outcome, which gives $\delta_T = 0.0785$.

\paragraph{Bootstrap and $p$-values}
Each of the 1,000 resamples draws identities or patients with replacement within each test fold, recomputes the fold metrics and their weights, and is shared by all conditions.
Summaries across CelebA pairs additionally resample the attributes that serve as $Z$, with all pairs that share one drawn as a block.
For the cross-group loss on CelebA, a resample is valid only if every pair still meets the cell-size rule of Section~\ref{sec:data}, which holds for 356 resamples over all pairs and for 460, 764, and 996 resamples in the low, middle, and high tertiles.
All resamples are valid for the other CelebA quantities and on CANDID-PTX.
The centered $p$-value is $(1 + b)/(B + 1)$, where $b$ counts valid resamples whose centered statistic lies beyond the boundary and $B$ is the number of valid resamples.
With $B = 1{,}000$, the smallest Holm-adjusted values are $12/1001 \approx 0.012$ on CelebA and $3/1001 \approx 0.003$ on CANDID-PTX, and we report results at these values as bounds.
Intervals are 95\% percentile intervals without multiplicity adjustment.

\paragraph{Design issues}
On CANDID-PTX, the attacker's validation data contain only 1 to 9 images with a tube but no pneumothorax per fold, so in every fold the conditional attacker fell back to selection by unconditional AUROC, as the plan specifies when any cell of $W \times Z$ in the validation data has fewer than 10 images.
On CelebA, the plan did not anticipate that re-applying the cell-size rule inside the bootstrap would invalidate resamples.
CANDID-PTX has no sealed partition.
Exploratory analyses with LEACE on all three encoders preceded the plan, and in one of them the MLP attack after LEACE fell below the imputation ceiling, which motivated H2$'$.
The results for the secondary CANDID-PTX encoders, for tertiles of $|\beta|$, and for single CelebA pairs, as well as all post-hoc analyses, are descriptive.

\section{Experimental Details}
\label{app:details}

\paragraph{Resolution}
DINOv2-base receives images of $518 \times 518$ pixels on CelebA and $224 \times 224$ pixels on CANDID-PTX.

\paragraph{Splits}
The outer folds are fixed identity (CelebA) or patient (CANDID-PTX) assignments that do not depend on any outcome.
Within each fold, the split into $A$ and $B$ and the internal validation splits use the seed $20261103$ plus the fold index, and a run stops if the realized proportions of $A$ and $B$ deviate from $80/20$ by over one percentage point.

\paragraph{Classifiers}
All logistic regressions minimize a class-balanced cross-entropy with an $\ell_2$ penalty $\lVert\mathbf{w}\rVert_2^2/(2Cn)$ on the coefficients but not the intercept, in float64 with L-BFGS (strong Wolfe line search, at most 1,000 iterations, gradient tolerance $10^{-10}$).
The attackers use $C = 1$, and the linear utility probe selects $C \in \{0.1, 1, 10\}$ by AUROC on $A_{\mathrm{val}}$.
The MLP attacker has two hidden layers of 256 and 64 ReLU units with dropout $0.2$ and is trained with AdamW (learning rate $10^{-3}$, weight decay $10^{-4}$, batch size 2,048) on a class-balanced cross-entropy for at most 200 epochs, with early stopping on the AUROC on $B_{\mathrm{val}}$ (patience 15) and the best epoch restored.
Its predicted probabilities are averaged over three seeds.
The MLP utility probe has one hidden layer of 256 ReLU units and is trained with a single seed in the same way for at most 120 epochs, with early stopping on $A_{\mathrm{val}}$ (patience 12), under two candidate settings (dropout 0 with learning rate $10^{-3}$, and dropout $0.1$ with learning rate $3\times10^{-4}$), of which the one with the higher AUROC on $A_{\mathrm{val}}$ is used.
Attacker inputs are standardized on $B_{\mathrm{train}}$ and probe inputs on $A_{\mathrm{train}}$.
No attacker hyperparameters are tuned beyond early stopping and the selection between logistic regression and the MLP.

\paragraph{Compute and code}
Each main-run job was allocated one NVIDIA H100 GPU, 20 vCPUs, and 200\,GiB of memory, with one job per CelebA encoder and one job for all CANDID-PTX encoders.
Summed over the 5,100 CelebA runs (four encoders, 85 pairs, five folds, three conditions), fitting the eraser and training and evaluating the probes and attackers took 71.0 hours, or 50\,s per run on average, and the 45 CANDID-PTX runs took about 22 minutes, excluding data loading and job overhead.
The CelebA bootstrap and summaries ran on 40 vCPUs in 38 minutes.
Embedding extraction, a single forward pass per image, was not reliably timed.
\textbf{Code is available at\\ \url{https://anonymous.4open.science/r/TC-LEACE-04E1/}.}

\section{Additional Results}
\label{app:more}

\paragraph{Simulation check of Corollary~\ref{cor:xauc}}
We simulated Gaussian embeddings with $n = 40{,}000$ and $d = 64$ in 27 settings that vary the association between $W$ and $Z$ (correlation $0$, $0.4$, $0.8$), the angle between the embedding directions along which $W$ and $Z$ shift the mean ($90^\circ$, $60^\circ$, $30^\circ$), and the number of random directions removed after LEACE ($0$, $8$, $32$), fitting the eraser and the probe on one sample and evaluating on an independent one.
The guardedness prediction of the cross-group loss erred by $0.003$ on average and by at most $0.011$.
Heavy-tailed or skewed noise and interaction terms gave errors of $0.002$ to $0.004$.
With $n/d \approx 31$, closer to the ratio in CANDID-PTX, the prediction overestimated the loss by $0.017$ to $0.029$, and inserting the measured residual $\varepsilon$ reduced the error to $0.001$ to $0.003$, as in Section~\ref{sec:case}.

\paragraph{Secondary encoders (Table~\ref{tab:candid-secondary}), post-hoc analyses (Table~\ref{tab:posthoc}), and CelebA pairs (Table~\ref{tab:pairs})}

\begin{table*}[!tp]
\centering
\begin{minipage}[t]{0.485\textwidth}
\centering
\footnotesize
\setlength{\tabcolsep}{4pt}
\begin{threeparttable}
\caption{Secondary encoders on CANDID-PTX.}
\label{tab:candid-secondary}
\begin{tabular}{@{}lcc@{}}
\toprule
 & DINOv2 & BiomedCLIP \\
\midrule
\multicolumn{3}{@{}l}{\emph{After LEACE}} \\
Conditional AUROC & 0.671 & 0.648 \\
\quad MLP attacker & 0.723 & 0.731 \\
\quad logistic attacker & 0.608 & 0.588 \\
Uncond.\ $-$ cond. & 0.096 & 0.168 \\
Linear probe loss & 0.169 & 0.170 \\
\quad cross-group part & 0.115 & 0.110 \\
\quad guardedness prediction & 0.144 & 0.148 \\
\quad ex-ante prediction & 0.136 & 0.139 \\
\midrule
\multicolumn{3}{@{}l}{\emph{\textsc{TC-LEACE} minus LEACE}} \\
Conditional AUROC & $+0.025$ & $+0.023$ \\
 & {\scriptsize[0.003, 0.046]} & {\scriptsize[0.007, 0.038]} \\
\quad MLP attacker & $+0.010$ & $+0.004$ \\
 & {\scriptsize[$-$0.004, 0.023]} & {\scriptsize[$-$0.008, 0.016]} \\
\quad logistic attacker & $+0.020$ & $+0.018$ \\
 & {\scriptsize[0.008, 0.031]} & {\scriptsize[0.007, 0.029]} \\
Linear probe loss & $-0.128$ & $-0.126$ \\
$\mathrm{AUC}_{00}$, linear & $+0.063$ & $+0.069$ \\
\bottomrule
\end{tabular}
\begin{tablenotes}[flushleft]
\footnotesize
\item Attack AUROCs use the combined attacker unless noted, and its fallback selection (Appendix~\ref{app:plan}) picks logistic regression in some folds.
Brackets give 95\% bootstrap intervals.
For the combined and the logistic attacker, the intervals of the differences in conditional AUROC exclude zero, whereas for the MLP attacker they include zero, and all upper bounds lie below $\delta_T = 0.0785$.
Losses are relative to raw embeddings, and both predictions apply Corollary~\ref{cor:xauc} with $\varepsilon = 0$.
\end{tablenotes}
\end{threeparttable}
\end{minipage}\hfill
\begin{minipage}[t]{0.485\textwidth}
\centering
\footnotesize
\setlength{\tabcolsep}{3pt}
\begin{threeparttable}
\caption{Post-hoc analyses on CANDID-PTX.}
\label{tab:posthoc}
\begin{tabular}{@{}lccc@{}}
\toprule
 & RAD-DINO & DINOv2 & BiomedCLIP \\
\midrule
\multicolumn{4}{@{}l}{\emph{Logistic attacker within strata, $W=0$ / $W=1$}} \\
Main, LEACE & 0.608/0.604 & 0.619/0.596 & 0.600/0.576 \\
Main, \textsc{TC-LEACE} & 0.651/0.594 & 0.666/0.585 & 0.635/0.575 \\
Release fit, LEACE & 0.564/0.530 & 0.576/0.551 & 0.560/0.526 \\
Release fit, \textsc{TC-LEACE} & 0.502/0.483 & 0.539/0.507 & 0.497/0.496 \\
\midrule
\multicolumn{4}{@{}l}{\emph{MLP attacker, conditional AUROC}} \\
Main, LEACE & 0.771 & 0.723 & 0.731 \\
Release fit, LEACE & 0.740 & 0.697 & 0.721 \\
Release fit, \textsc{TC-LEACE} & 0.655 & 0.660 & 0.662 \\
\midrule
\multicolumn{4}{@{}l}{\emph{Orthogonalization, MLP attacker, main setting}} \\
Conditional AUROC & 0.944 & 0.880 & 0.946 \\
\quad raw embeddings & 0.877 & 0.814 & 0.803 \\
\quad difference & $+0.067$ & $+0.066$ & $+0.142$ \\
 & {\scriptsize[0.045, 0.091]} & {\scriptsize[0.048, 0.086]} & {\scriptsize[0.119, 0.166]} \\
\bottomrule
\end{tabular}
\begin{tablenotes}[flushleft]
\footnotesize
\item Main denotes the main setting, in which the eraser is fitted on $A_{\mathrm{train}}$ and applied to new patients.
Release fit denotes the setting in which the eraser is fitted on the embeddings it releases ($A_{\mathrm{train}}$, $A_{\mathrm{val}}$, and the test fold) but not on the attacker's data $B$.
We report only attack results for the release fit, because the utility probes then train inside the fitting data.
Orthogonalization follows~\cite{weber2025orthogonalization}, and brackets give 95\% bootstrap intervals.
For retrained linear readers, SPLINCE differs from LEACE by at most $0.003$ in the main setting and $0.009$ in the release fit.
\end{tablenotes}
\end{threeparttable}
\end{minipage}

\vspace{1.2em}

{\scriptsize
\setlength{\tabcolsep}{1.5pt}
\caption{The 85 CelebA pairs $(W, Z)$ by tertile of $|\beta|$, sorted by $|\beta|$ within each tertile, with $\beta$ from the main experiment.}
\label{tab:pairs}
\begin{tabular}{@{}llr@{\hspace{5pt}}llr@{\hspace{5pt}}llr@{}}
\toprule
\multicolumn{3}{c}{High tertile} & \multicolumn{3}{c}{Middle tertile} & \multicolumn{3}{c}{Low tertile} \\
\cmidrule(lr){1-3}\cmidrule(lr){4-6}\cmidrule(lr){7-9}
$W$ & $Z$ & $\beta$ & $W$ & $Z$ & $\beta$ & $W$ & $Z$ & $\beta$ \\
\midrule
Smiling & High Cheekbones & 0.684 & Rosy Cheeks & Wearing Necklace & 0.102 & Receding Hairline & Rosy Cheeks & $-$0.026 \\
Heavy Makeup & Wearing Earrings & 0.439 & Wearing Earrings & Blurry & $-$0.102 & Straight Hair & Big Nose & $-$0.025 \\
Wearing Lipstick & Blond Hair & 0.404 & Mouth Slightly Open & Male & $-$0.099 & Wearing Earrings & Receding Hairline & 0.024 \\
Arched Eyebrows & Heavy Makeup & 0.397 & Black Hair & Sideburns & 0.090 & Narrow Eyes & Heavy Makeup & $-$0.024 \\
High Cheekbones & Wearing Earrings & 0.299 & Narrow Eyes & Big Lips & 0.086 & Big Lips & Goatee & 0.023 \\
Heavy Makeup & Young & 0.285 & Young & Wavy Hair & 0.084 & Young & 5 o'Clock Shadow & 0.022 \\
High Cheekbones & Wearing Lipstick & 0.281 & Eyeglasses & Big Nose & 0.083 & Young & Wearing Necklace & 0.019 \\
High Cheekbones & Heavy Makeup & 0.275 & Oval Face & Wearing Necklace & $-$0.082 & Receding Hairline & Narrow Eyes & 0.018 \\
Big Nose & Wearing Lipstick & $-$0.259 & Straight Hair & Wearing Earrings & $-$0.080 & Young & High Cheekbones & $-$0.018 \\
Bags Under Eyes & Heavy Makeup & $-$0.244 & Blurry & Attractive & $-$0.079 & Receding Hairline & Smiling & 0.016 \\
Wearing Lipstick & Big Lips & 0.233 & Narrow Eyes & Mouth Slightly Open & 0.074 & Straight Hair & Pointy Nose & $-$0.016 \\
Wearing Necktie & Chubby & 0.214 & Big Lips & Big Nose & 0.071 & Bushy Eyebrows & Big Lips & 0.015 \\
Wavy Hair & Attractive & 0.201 & Straight Hair & Goatee & $-$0.068 & Narrow Eyes & Bangs & 0.014 \\
Big Nose & Blond Hair & $-$0.194 & Wavy Hair & Smiling & 0.068 & Blurry & Eyeglasses & 0.014 \\
Brown Hair & Blond Hair & $-$0.190 & Goatee & Wearing Necktie & 0.066 & Black Hair & Narrow Eyes & $-$0.014 \\
Oval Face & Smiling & 0.187 & Black Hair & Big Lips & 0.062 & Narrow Eyes & Wearing Necktie & 0.012 \\
Chubby & Big Nose & 0.172 & High Cheekbones & Receding Hairline & 0.060 & Wearing Necktie & Mouth Slightly Open & $-$0.012 \\
Young & Sideburns & $-$0.156 & Attractive & Straight Hair & 0.048 & Eyeglasses & Black Hair & $-$0.010 \\
Male & Black Hair & 0.143 & Brown Hair & Oval Face & 0.043 & Blurry & Wavy Hair & $-$0.009 \\
Brown Hair & Eyeglasses & $-$0.128 & Rosy Cheeks & Big Lips & 0.043 & Narrow Eyes & Wearing Earrings & 0.009 \\
Bags Under Eyes & Bushy Eyebrows & 0.125 & 5 o'Clock Shadow & Straight Hair & 0.042 & High Cheekbones & Black Hair & 0.008 \\
Bags Under Eyes & Blond Hair & $-$0.123 & Eyeglasses & Bangs & $-$0.042 & Narrow Eyes & Black Hair & $-$0.008 \\
Bangs & Male & $-$0.119 & Big Lips & Blond Hair & 0.042 & Brown Hair & Wearing Necklace & $-$0.006 \\
Young & Black Hair & 0.119 & Pointy Nose & Smiling & 0.037 & Black Hair & Arched Eyebrows & $-$0.004 \\
Smiling & Blurry & $-$0.118 & Oval Face & Wavy Hair & 0.036 & Wearing Necklace & Brown Hair & $-$0.004 \\
Bangs & Wearing Lipstick & 0.117 & Narrow Eyes & Wearing Necklace & 0.031 & Bangs & Oval Face & $-$0.003 \\
5 o'Clock Shadow & Big Nose & 0.114 & Straight Hair & Sideburns & $-$0.029 & Chubby & Big Lips & 0.003 \\
Brown Hair & Wearing Necktie & $-$0.112 & Wearing Necklace & Big Nose & $-$0.026 & Narrow Eyes & Sideburns & 0.000 \\
 & &  &  & &  & Smiling & Bushy Eyebrows & 0.000 \\
\bottomrule
\end{tabular}
}
\end{table*}
\end{document}